%% file: ms.tex
\documentclass[12pt,preprint]{aastex}
\usepackage{psfig,natbib}
 
\newcommand{\feka}{Fe K$\alpha$}
\newcommand{\chandra}{{\it Chandra}}

\newcommand{\xmm}{{\it XMM--Newton}}
\newcommand{\asca}{{\it ASCA}}
\newcommand{\hst}{{\it HST}}
\newcommand{\sax}{{\it BeppoSAX}}
\newcommand{\lum}{erg s$^{-1}$}

\newcommand{\nh}{cm$^{-2}$}
  
\shorttitle{An X--ray View of WLRGs/LINERs}
\shortauthors{Rinn et al.}

\begin{document}  
  
\title{An X--ray View of WLRGs/LINERs}

\author{Alexander S. Rinn, Rita M. Sambruna, and Mario Gliozzi}
\affil{George Mason University, Dept. of Physics and Astronomy, MS 3F3, 
4400 University Drive, Fairfax, VA 22030}

\begin{abstract}
We present X--ray observations of nine Weak-Line Radio Galaxies (WLRGs),
optically classified as confirmed or possible Low Ionization Nuclear
Emission--line Regions (LINERs).  The data were taken from the
\chandra, \xmm, and \sax\ archives.  The \chandra\ images typically show 
complex
X-ray morphologies, with hard (2--10 keV) point sources embedded in diffuse 
soft (0.3--2.0 keV) emission in all cases except 1246--41 (NGC 4696), 
where only diffuse emission
is detected on the scale of the cluster, and 0334--01 (3C 15), where 
only a point source is detected.   The nuclear X--ray spectra
are well fitted at hard energies by an absorbed power
law, with a wide range of photon indices, $\Gamma=1.5-2.7$. Excess
absorption over the Galactic value is detected in 6/9 sources, with
column densities N$_H \approx 10^{21-22}$ \nh.  A thermal component is
required at softer energies, in agreement with the results of the
spatial analysis. We find that there is no correlation between the core
X--ray luminosity and the radio core dominance parameter, suggesting that the
bulk of the core X-ray emission is not beamed, but rather is isotropic and 
thus likely
related to the accretion flow. In an attempt to constrain the nature of the
accretion flow, we calculate the ratios of bolometric to Eddington
luminosities $L_{bol}/L_{Edd}$, and the radiative efficiency $\eta$
based on the Bondi accretion rates. We find that 
$L_{bol}/L_{Edd} \sim 10^{-4}-10^{-6}$ and $\eta \sim 10^{-2}-10^{-6}$   
for all the objects in
our sample, suggesting radiatively inefficient accretion flows.
 
\end{abstract}

   
\section{Introduction}

The recent discovery of \hst\ and ground--based observations that most
nearby galaxies harbor supermassive black holes \citep{kor95,ric98},
along with independent evidence that weak nuclear activity is very
common in both spiral and elliptical galaxies
\citep{hec80,sad89,tad98}, affirms the fundamental link between
``normal'' and active galaxies.  According to the Palomar survey
\citep{ho97}, over 40\% of nearby galaxies ($z\approx0$) contain
low--power Active Galactic Nuclei (AGN), with typical bolometric
luminosities lower than $10^{44}$ \lum.  Most of these, based on their
low--ionization spectra, are classified as Low Ionization Nuclear
Emission--line Regions (LINERs), or transition objects.  Since
accreting supermassive black holes are widely held to be responsible
for the nuclear activity in AGN, one naturally wonders why such
intense nuclear activity is limited to a minority of galaxies.  A
possible explanation is that obscuring material along the line of
sight conceals the nuclear activity.  Alternatively, there could be
inefficient accretion onto the black hole due either to a low
accretion rate (the AGN--starved scenario), or to some kind of
radiatively--inefficient accretion flow \citep[RIAF; see
e.g.][]{qua03}.  Weak AGN, such as those hosted by nearby low--power
radio galaxies in giant ellipticals, may represent the link between
powerful AGN and ``normal'' galaxies.

One class of objects that may represent this link are Weak--Line Radio
Galaxies \citep[WLRGs;][]{tad98}, which are classified by
[O$_{~\rm{III}}$]$\lambda$5007 luminosities more than 10 times lower
than those of the general population at similar redshifts.  The
weakness of the [O$_{~\rm{III}}$]$\lambda$5007 emission line in WLRGs
raises the possibility that these objects may be related to LINERs;
indeed, there is evidence that WLRGs and LINERs are intimately related
\citep{lew03}.  If all LINERs can be regarded as true AGN, then they
would constitute approximately 75\% of all AGN \citep{ho97}.  Put in this 
light,
WLRGs/LINERs could prove essential to understanding the link between
``normal'' and active galaxies.

X--rays provide us with a powerful tool by which we can achieve progress 
in our
understanding of the nature of accretion in low-power AGN, because
they originate in the innermost, hottest parts of the accretion
flow. In addition, soft X--rays map the distribution of the
circumnuclear gas around the nuclei, allowing us to quantify the fuel
reservoir on the nucleus.

In Section 2 we describe our sample of WLRGs/LINERs.  In Section 3 we
describe the observations and data reduction.  Section 4 contains the
results of the spatial analysis based on \chandra\ observations (4.1),
and the results of the temporal and spectral analysis (Sections 4.2 and 4.3,
respectively), both of which are based on \chandra, \xmm, and \sax\
observations.  In Section 5 we discuss the obscuration and nature of
accretion in WLRGs/LINERs, as well as some comparisons with the results of
previous work.  Finally, in Section 6 we summarize our results.
Throughout the paper we use a Friedman cosmology with $H_0={\rm
75~km~s^{-1}}$ and $q_0=0.5$.

\section{The Sample}

Our sample consists of 9 out of the 24 WLRGs that comprise the Lewis, 
Eracleous, \& Sambruna
(2003) sample.  That sample was selected in order to
investigate the relationship between WLRGs and LINERs.  Our 9 sources
were selected based on two criteria: first, that Lewis et
al. classified the source as a LINER, as a possible--LINER, or that
there were conflicting classifications; second, that there existed
X--ray observations for the sources made by either \chandra, \xmm, or
\sax. 

Table~\ref{tbl-1} lists some characteristics of the sources in our
sample.  It should be noted that all of the sources have redshifts
$z<$0.1.  The black hole masses were calculated using the relationship
$M_{BH}=1.3\times10^8(\sigma/200{\rm km~s^{-1}})^{4.58}$
\citep{fer00,mer03} and using velocity dispersions from the on line
catalog HyperLeda (http://foehn.univ-lyon1.fr/hypercat/\-search.html),
unless otherwise specified in the table.  These calculations were
consistent with those found in the literature.  Also, there is only
one source that is classified as an FRII (1637--77); the rest of the
sources are classified as either FRIs or transition objects FRI/II.
Finally, we list how Lewis et al. (2003) classified the sources: 6 of
the sources are confirmed LINERs, 2 are possible LINERs, and 1 is
conflicting.

\section{Observations and Data Reduction}

Table~\ref{tbl-2} summarizes the X--ray observations of the sources
comprising our sample.  We used 7 \chandra\ observations, 4 \xmm\
observations, and 1 \sax\ observation.

\subsection{\chandra}

All of the \chandra\ observations were made using the ACIS--S
detector.  Pile--up was not a problem in any of the observations.  The
data reduction was performed with CIAO v. 2.3 and CALDB v. 2.18.  The
data were reprocessed with {\tt acis\_process\_events} and screened to
exclude periods of background flaring.  In order to take into account
the continuous degradation of the ACIS quantum efficiency (QE) due to
molecular contamination of the optical blocking filters, we applied
{\tt ACISABS} to the ancillary response function (ARF) file.

The spectra were rebinned (with the exception of 1251--12) such that
each spectral bin contained at least 20 counts in order to apply
$\chi^2$ minimization, and fitted using the XSPEC v. 11.2 software
package \citep{arn96}.  Due to the very low number of counts, we did
not rebin the spectrum of 1251--12, and instead of the $\chi^2$
statistics, we used the C--statistics when fitting the spectrum.  The
quoted errors on the derived best--fit model parameters correspond to
a 90\% confidence level.  The latest publicly available responses were
used.  The same procedure was applied to the \xmm\ and \sax\ spectral
analysis.

\subsection{\xmm}

The \xmm\ analysis was based on the EPIC pn data, since they have the
highest signal--to--noise ratio among the EPIC cameras.  For 0915--11
and 1637--77 the EPIC pn operated in full-frame extended mode with a
thin filter.  For 1216+06 and 1246--41 the EPIC pn operated in
full--frame mode with a thin filter.  The recorded events were
reprocessed and screened with the latest available release of the
\xmm\ Science Analysis Software (SAS 5.4) to remove known hot pixels
and other data flagged as bad---only data with {\tt FLAG=0} were used.
We investigated the full-field light curves to detect periods of
background flaring; any such events were screened.  In 0915--11,
1216+06, and 1637--77 background data were extracted from a
source--free circular region on the same chip containing the source.
In 1246--41 background data were extracted in an annular region
centered about the nuclear extraction region; this was done in an
attempt to take into account the cluster in which this source lies.
There are no signs of pile-up in any of the sources.

\subsection{\sax}

\sax\ observed 0305+03 (3C 40) with the Narrow Field Instruments, LECS 
(0.1--4.5 keV) and MECS (1.3--10.0 keV) on 1997 January 07, with effective 
exposures of 8.7 ks and 19.4 ks respectively.  The LECS and MECS are both 
imaging instruments.  The observations were performed with 2 active MECS 
units.  Standard data reduction techniques were employed, following the 
prescription given by Fiore et al. (1999).  LECS and MECS spectra and light 
curves were extracted from regions with radii of 8' and 4' respectively, 
in order to maximize the accumulated counts at both low and high energies.  
Background spectra were extracted from high Galactic latitude ``blank'' 
fields.  The background subtracted count rates are 
$(3.7\pm{0.3}) \times 10^{-2}$ cnts s$^{-1}$ for the LECS in the 0.1--4.5 
keV band, and $(3.5\pm{0.1}) \times 10^{-2}$ cnts s$^{-1}$ for the MECS 
in the 2--10 keV energy band.

\section{Results}

\subsection{Spatial Analysis}

\chandra's subarcsecond spatial resolution provides us with the opportunity 
to disentangle the different X--ray components in a given source.  In 
particular, the detection of a hard X--ray nuclear point source lends 
support to the claim that a source houses an AGN.  Whatever the results of 
the spatial analysis are, they help one to better constrain and interpret the 
spectral data associated with the source.  

\begin{figure}
\begin{center}

\includegraphics[height=4.8cm,width=5.3cm]{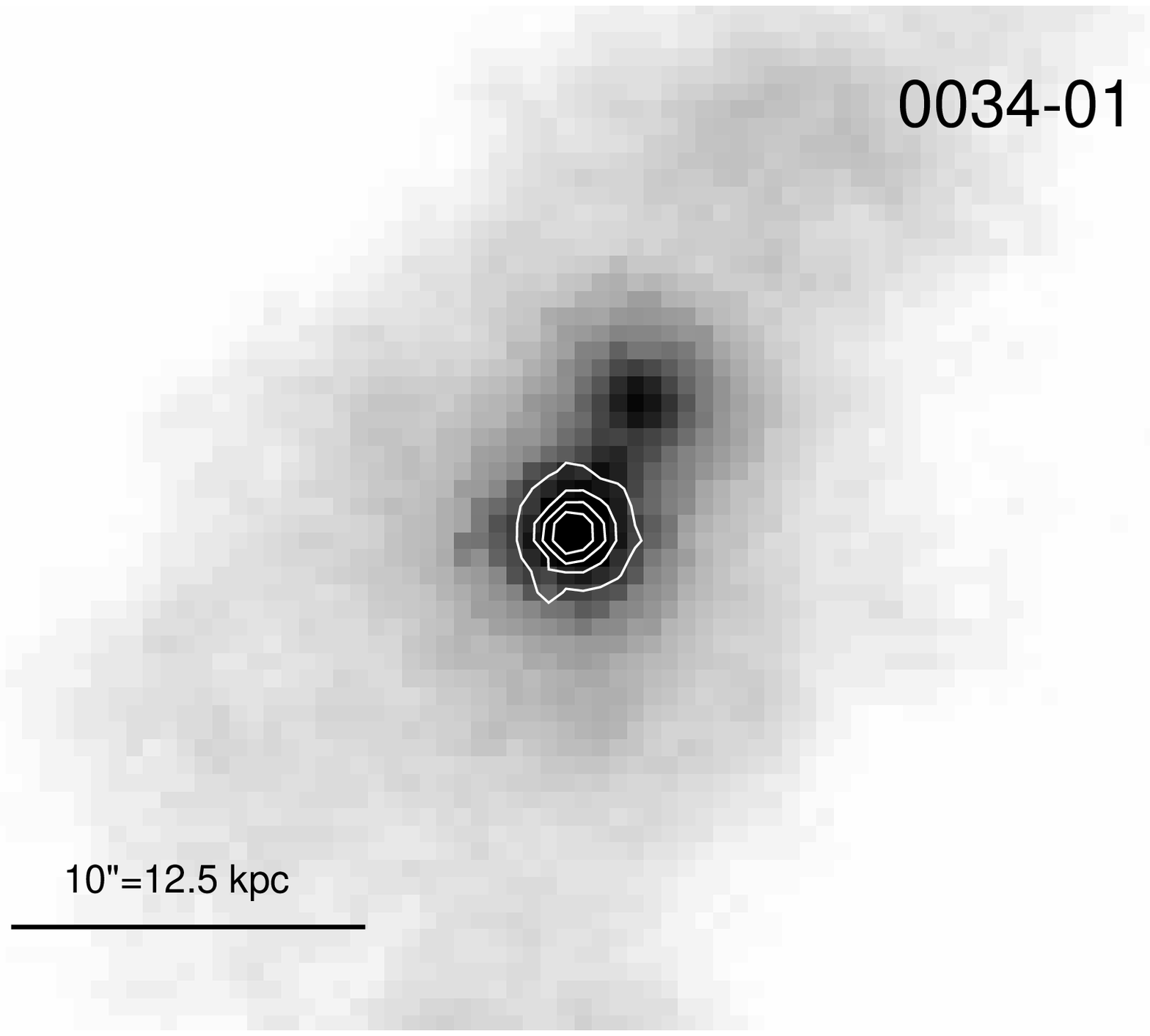}\includegraphics[height=4.8cm,width=5.3cm]{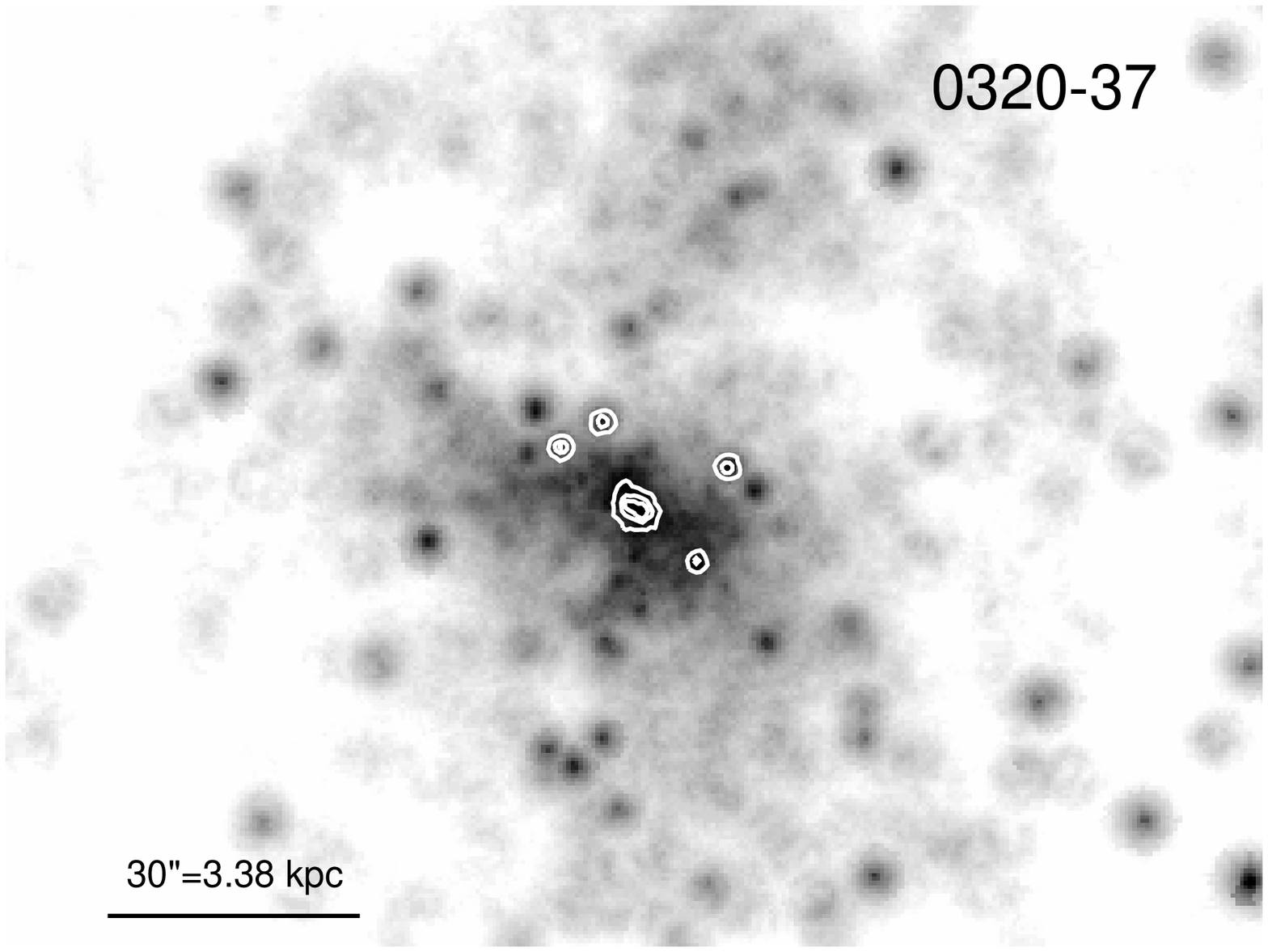}\includegraphics[height=4.8cm,width=5.3cm]{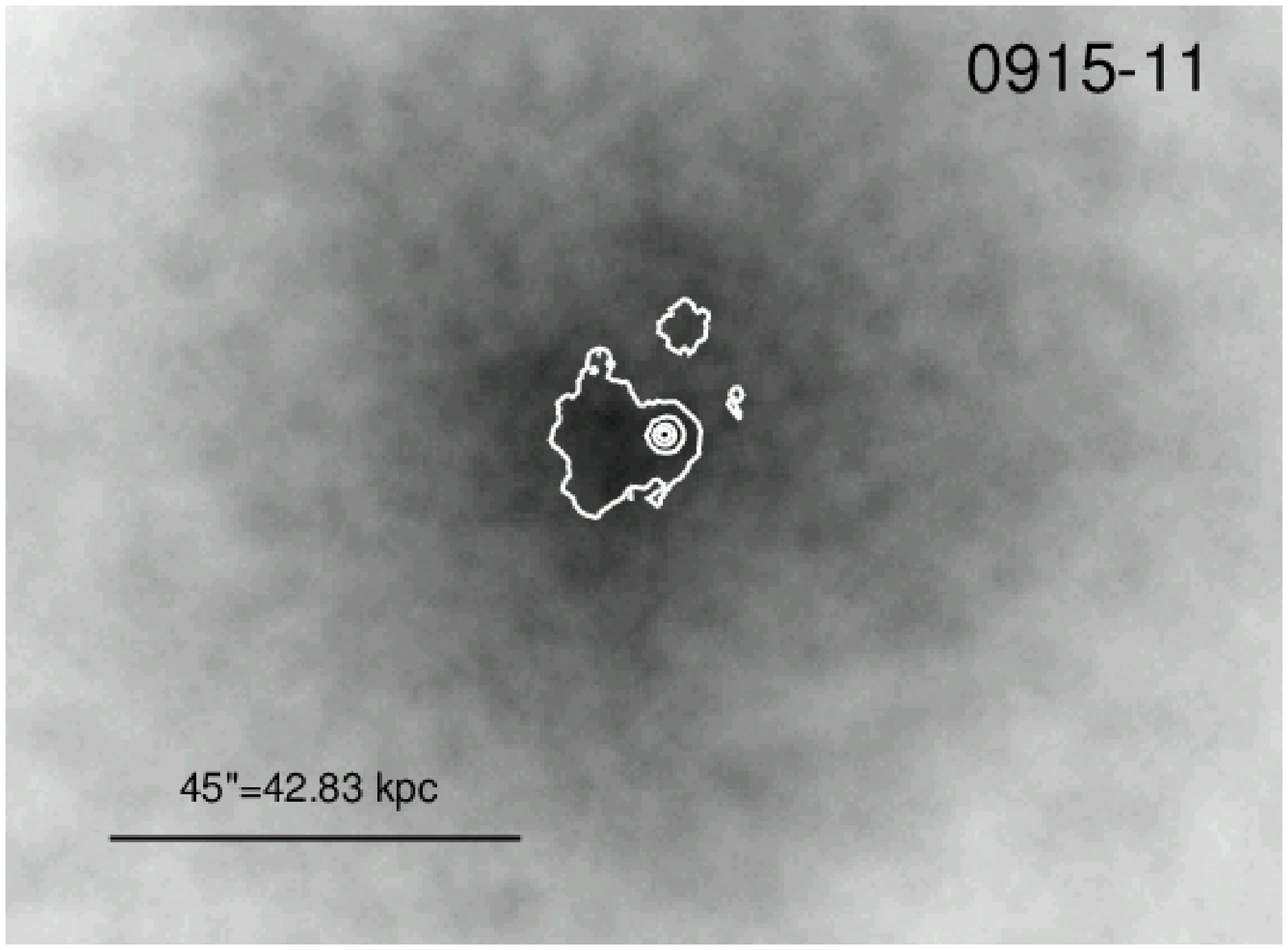}
\vskip 0.3 cm
\includegraphics[height=4.8cm,width=5.3cm]{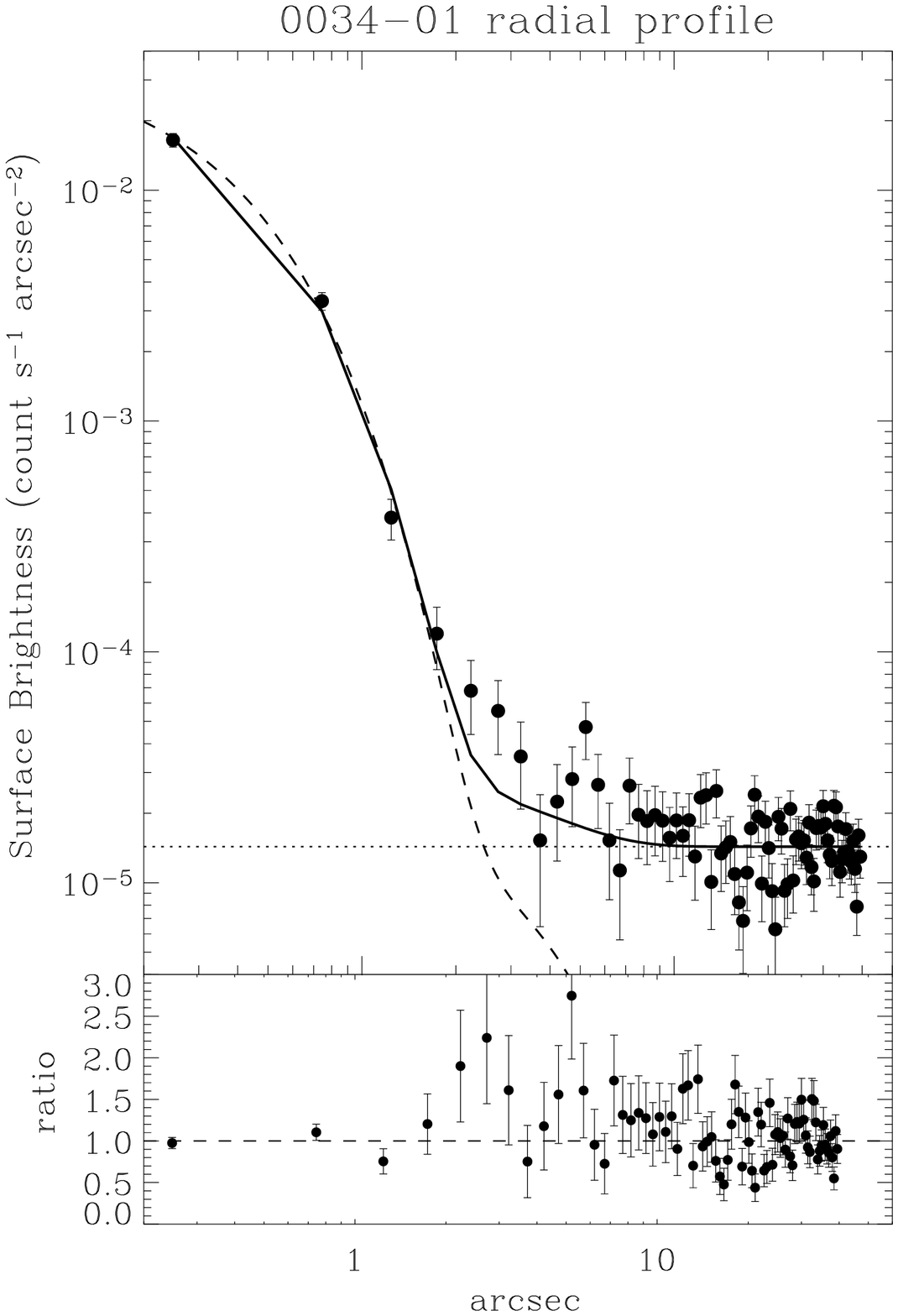}\includegraphics[height=4.8cm,width=5.3cm]{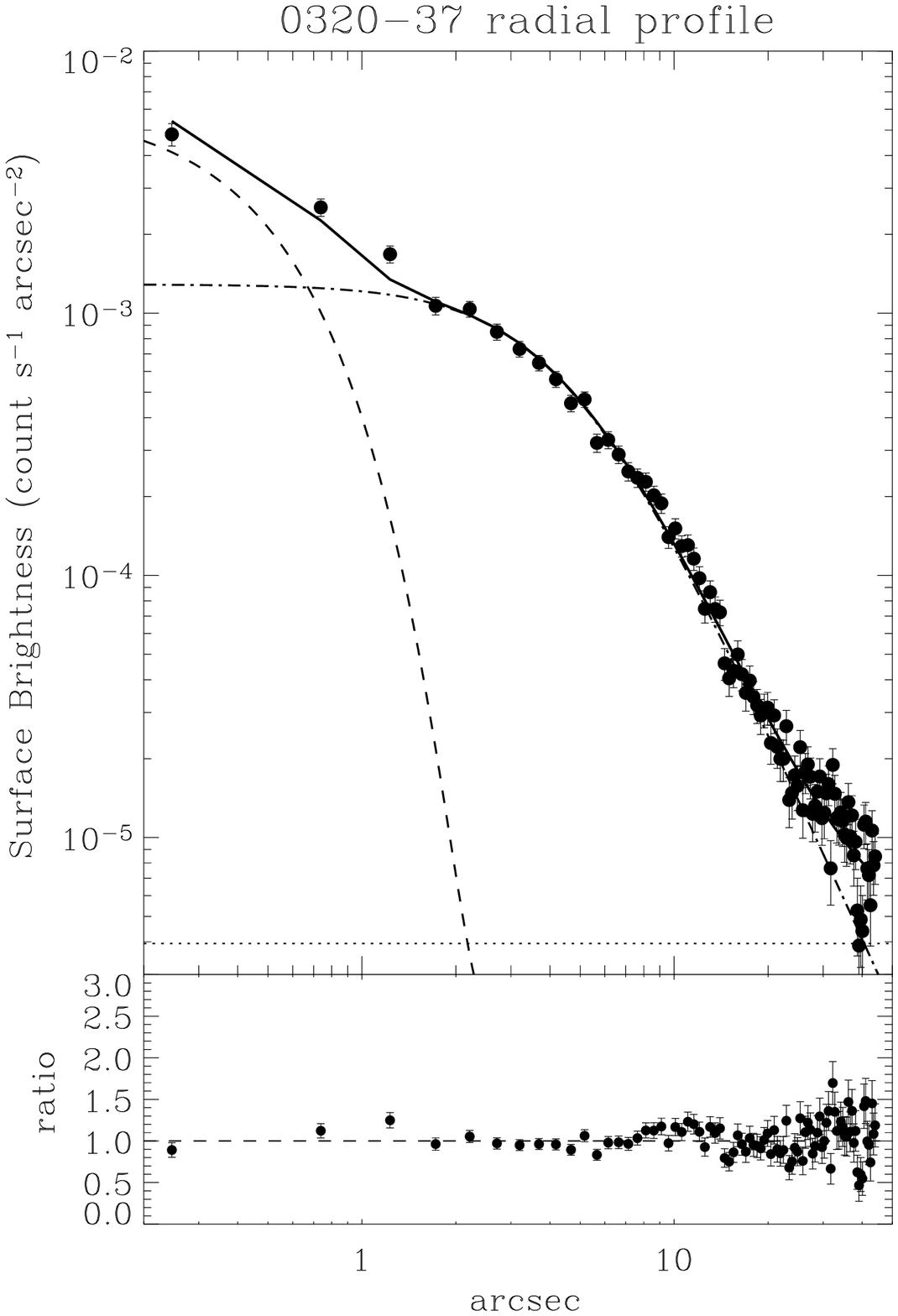}\includegraphics[height=4.8cm,width=5.3cm]{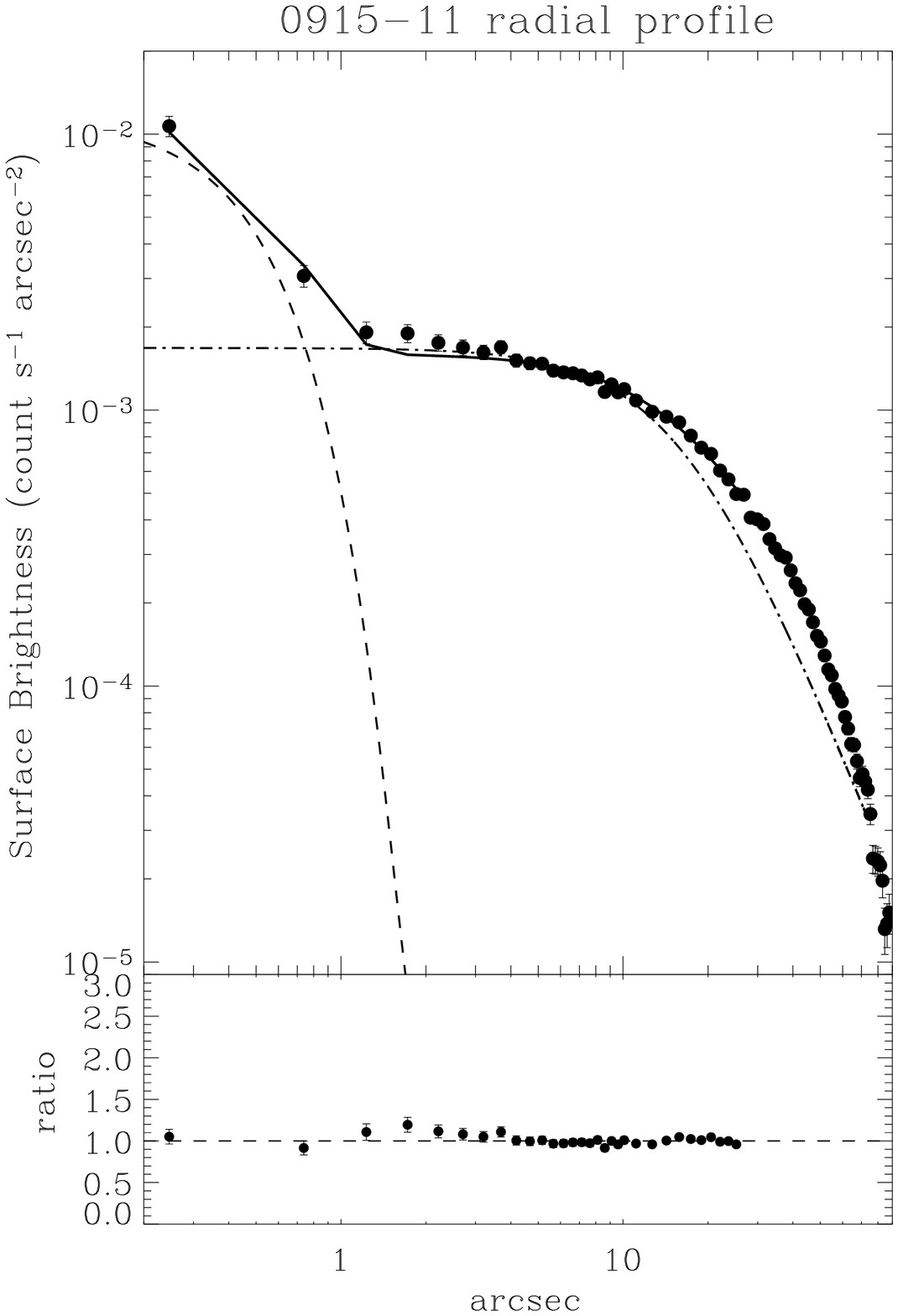}
\vskip 0.5 cm
\includegraphics[height=4.8cm,width=5.3cm]{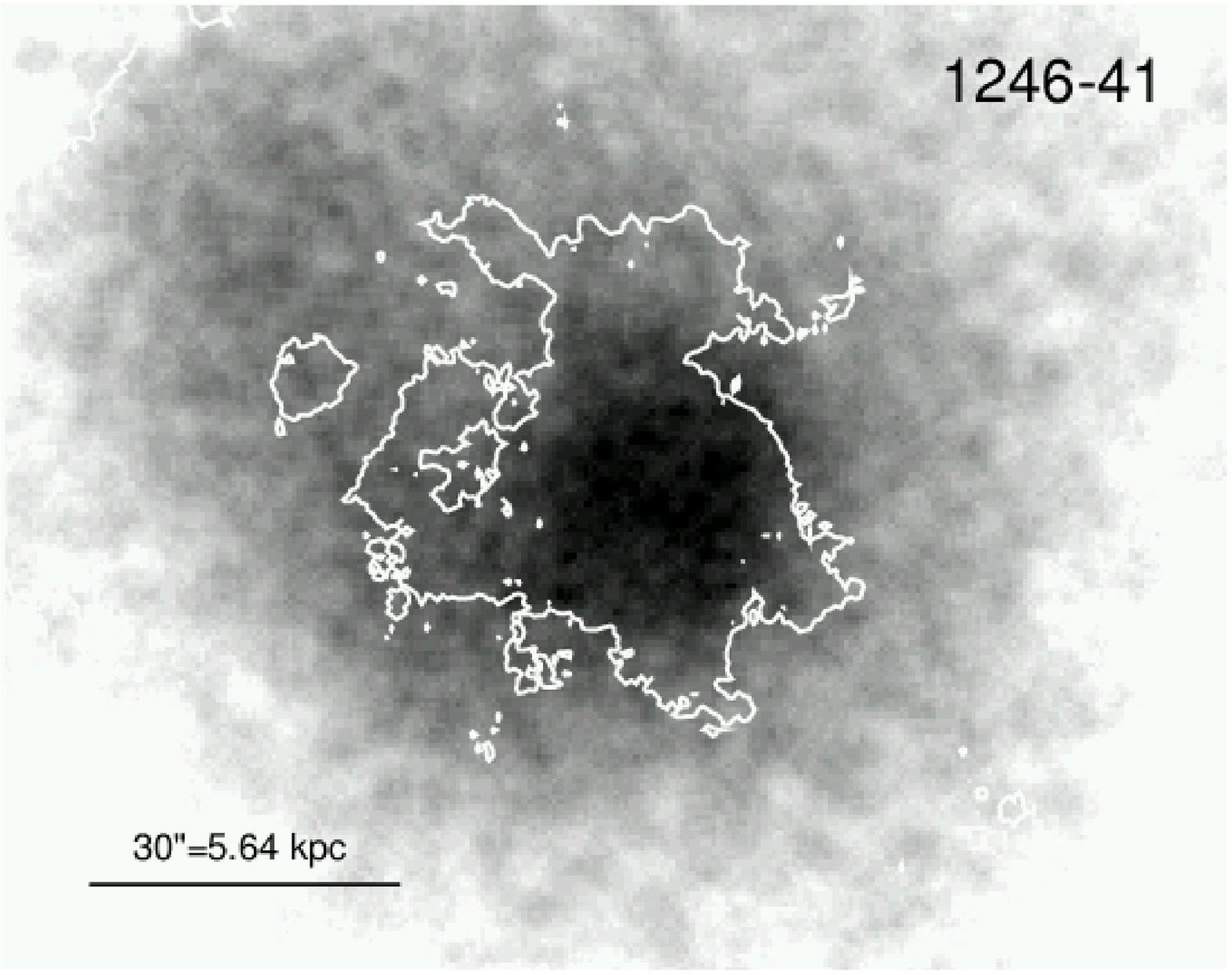}\includegraphics[height=4.8cm,width=5.3cm]{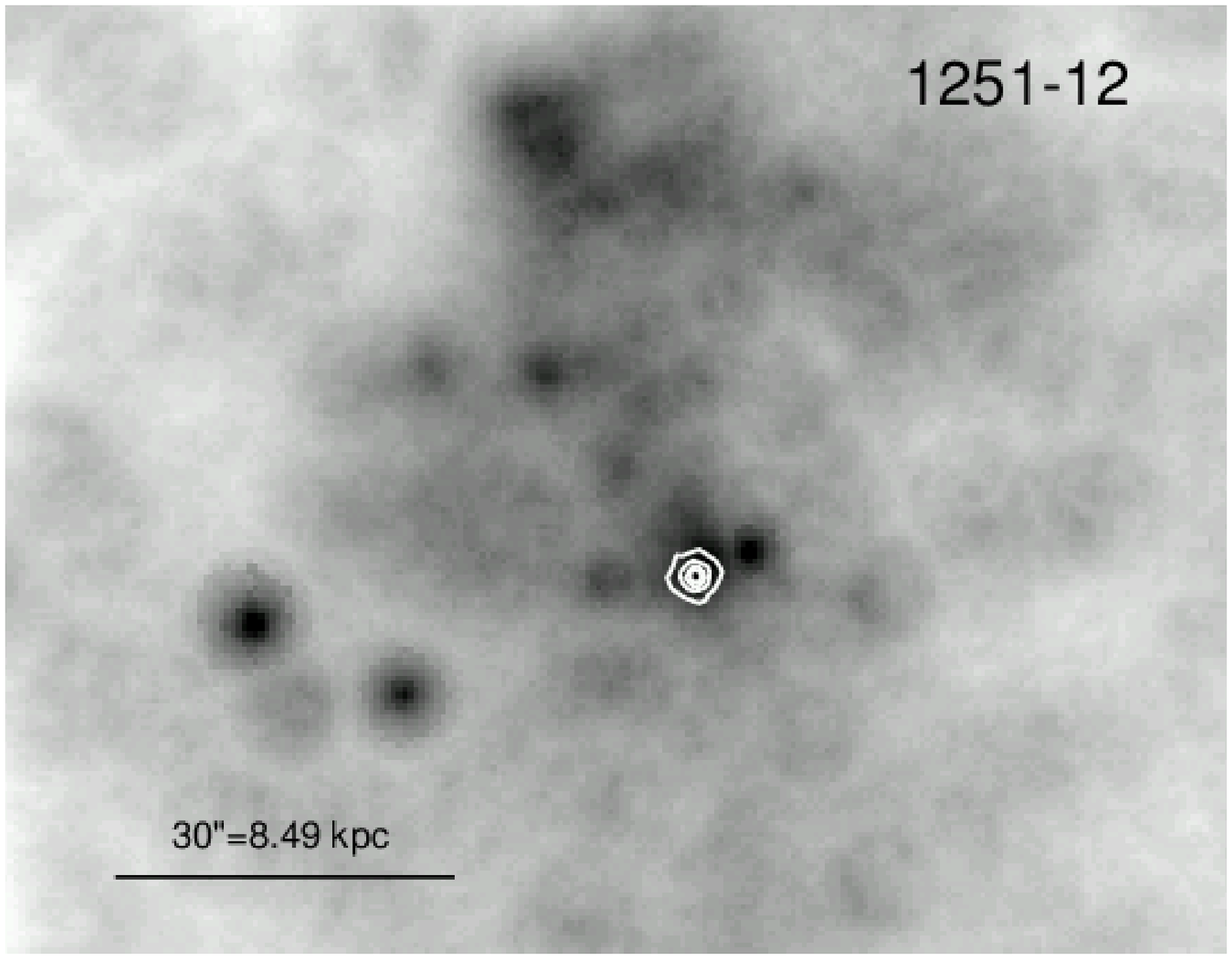}\includegraphics[height=4.8cm,width=5.3cm]{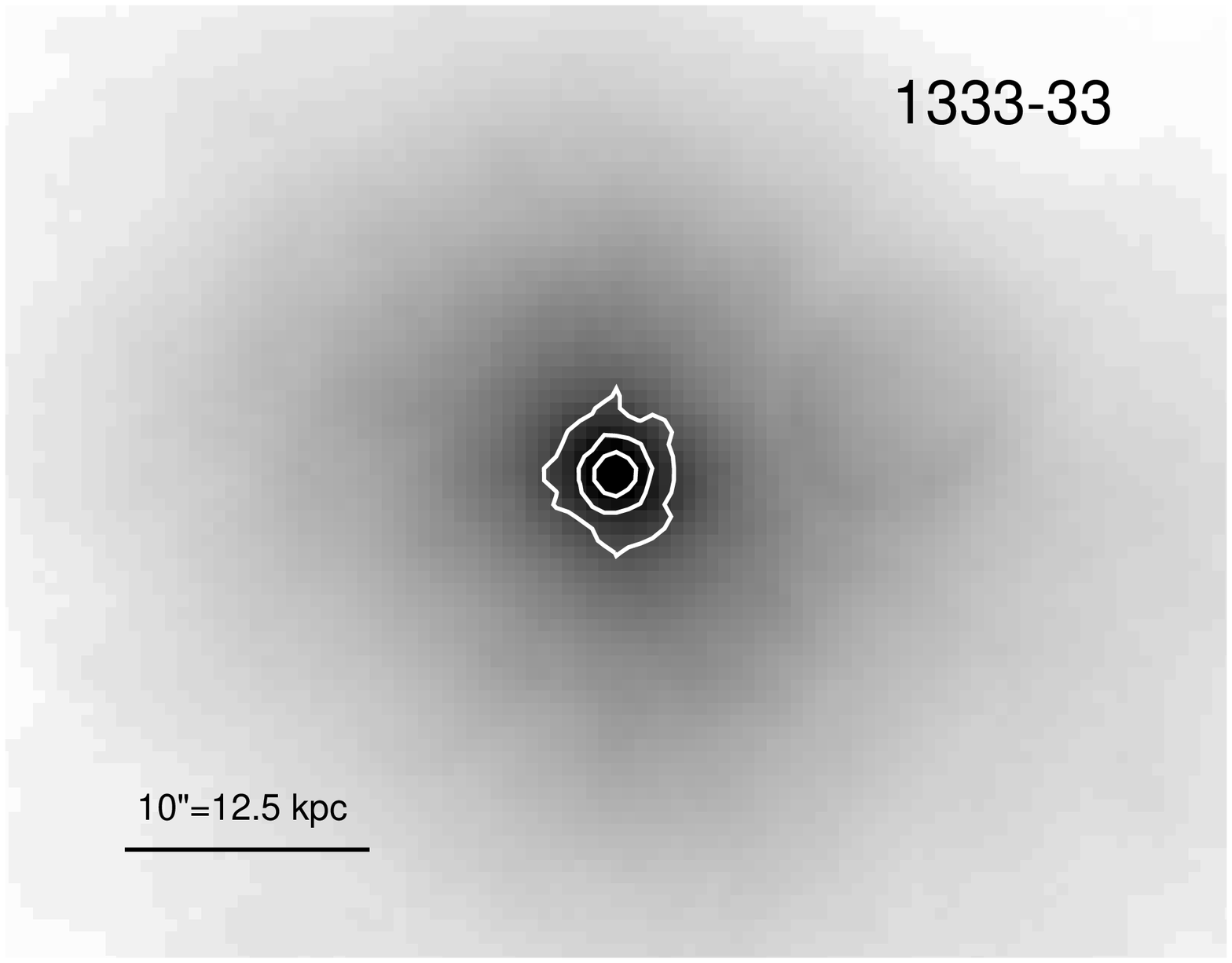}
\vskip 0.3 cm
\includegraphics[height=4.8cm,width=5.3cm]{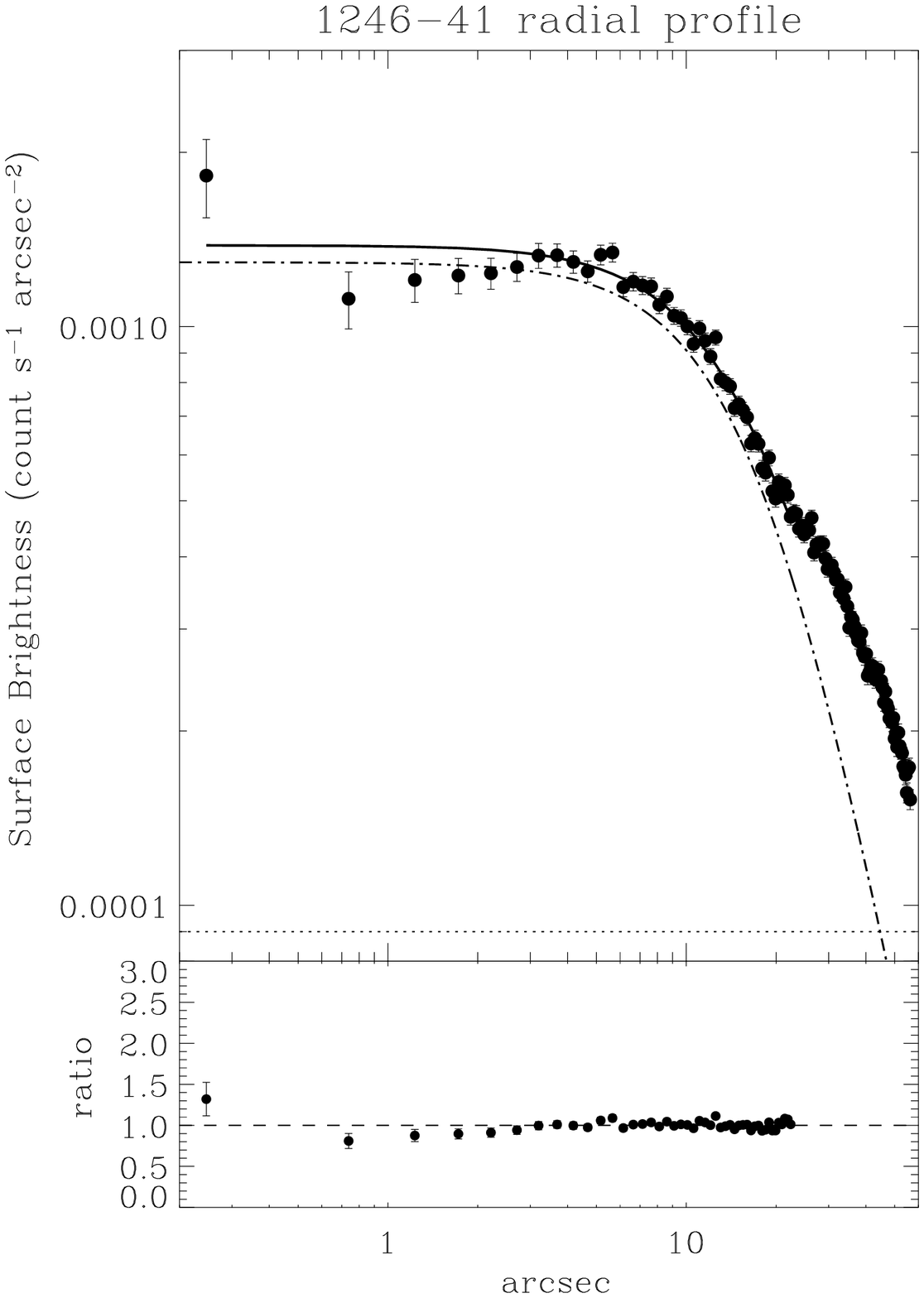}\includegraphics[height=4.8cm,width=5.3cm]{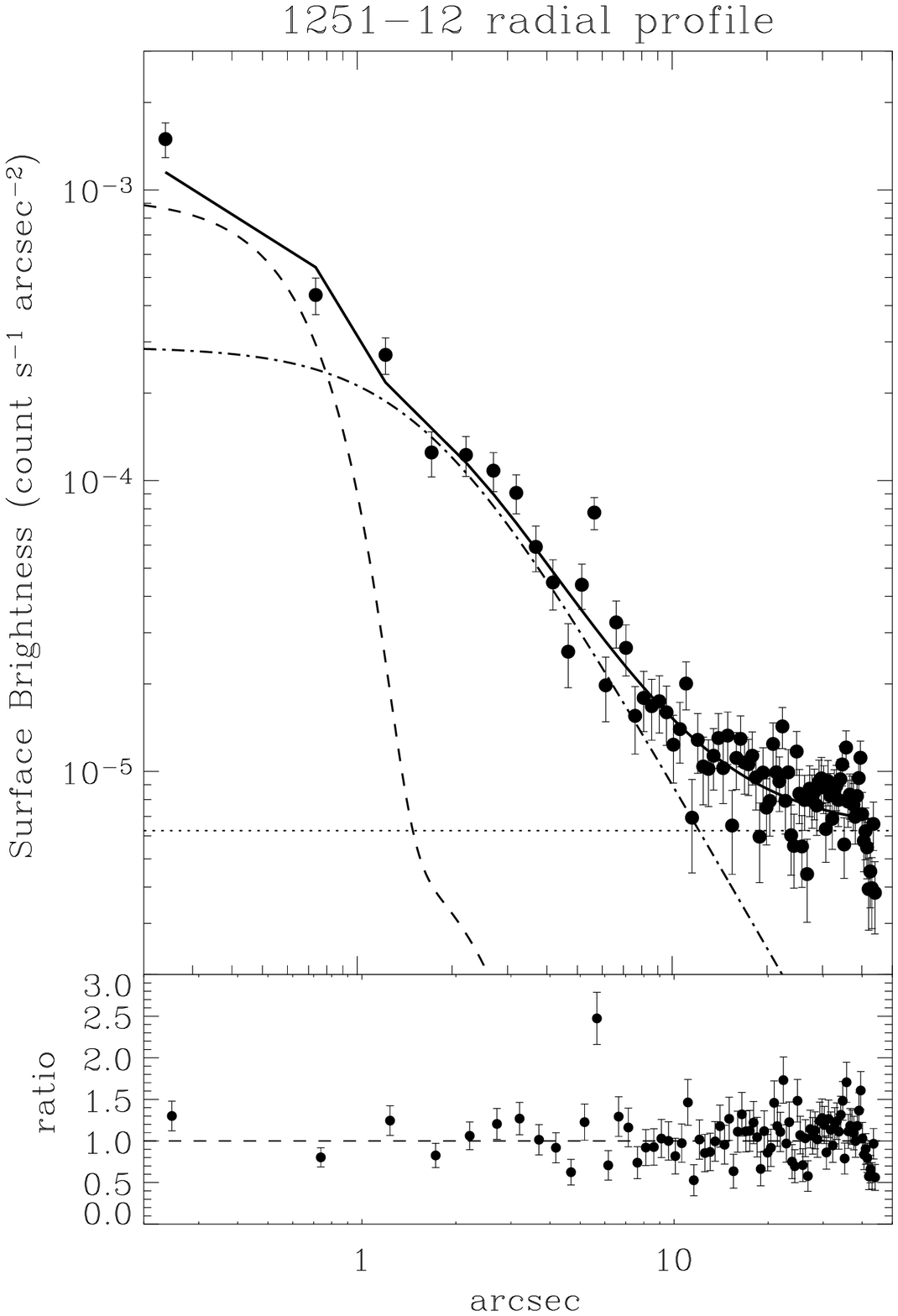}\includegraphics[height=4.8cm,width=5.3cm]{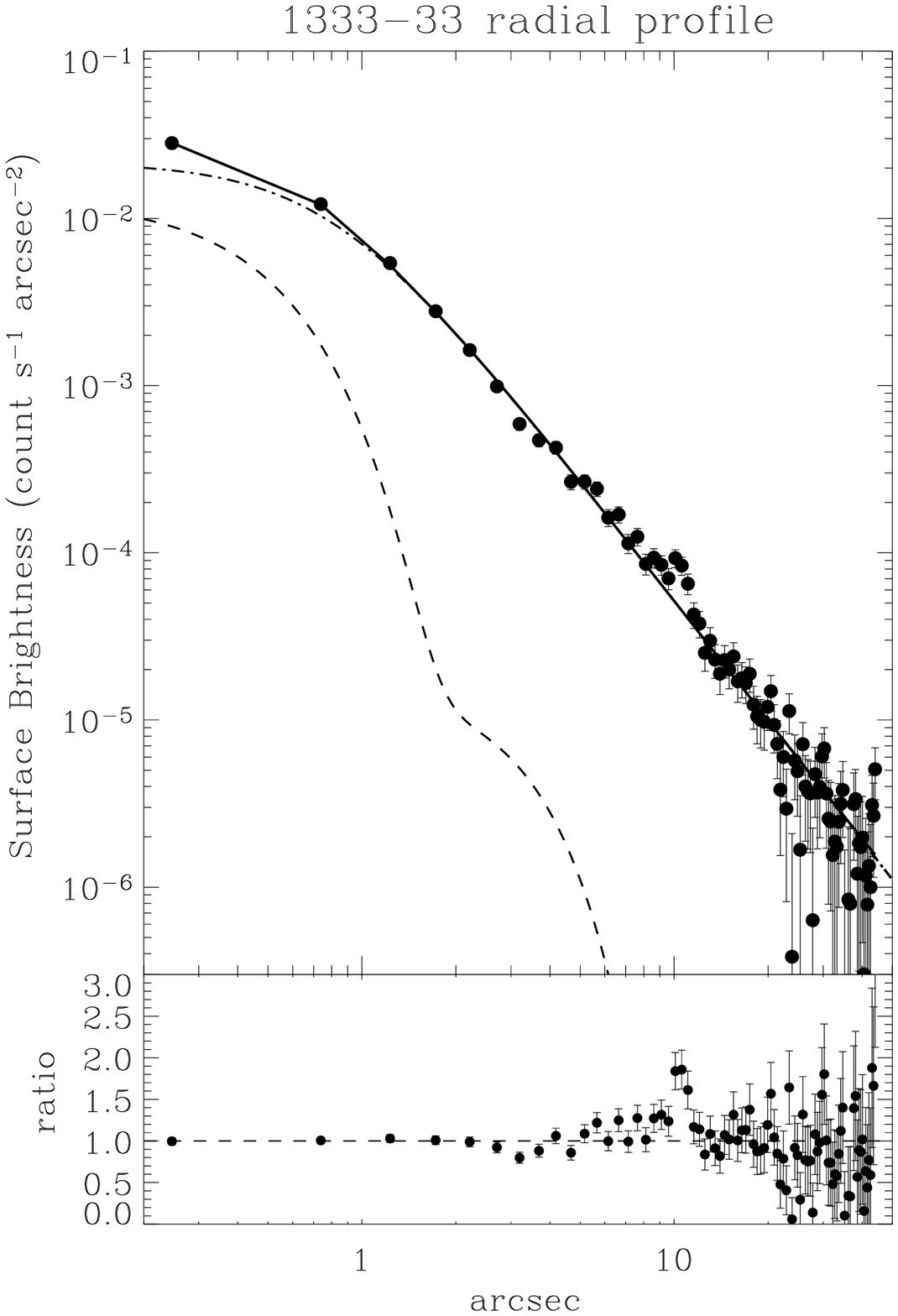}

\end{center}
\caption{Rows 1 and 3 show the adaptively smoothed \chandra\ images in the 0.3--10 keV range, with hard contours (2--10 keV) overlaid.  Rows 2 and 4 show the corresponding surface brightness profiles with best--fit (solid line), PSF (dashed line), and $\beta$--model (dot--dashed line) overlaid.
\label{spatial}}
\end{figure}

For each of the objects in our sample observed by \chandra\
\citep[except 1216+06, for which a more detailed spatial analysis is
given in][]{gli03}, Fig.~\ref{spatial} shows the adaptively smoothed
images in the energy band 0.3--10 keV with hard X--ray contours
superimposed (2--10 keV); also shown are the surface brightness
profiles.  The images were smoothed using {\tt fadapt} from the {\tt
ftools} software package.  To produce the surface brightness profiles,
we first used the source energy spectrum to derive the appropriate
Point Spread Function (PSF) with ChaRT
(http://cxc.harvard.edu/soft/ChaRT/cgi-bin/www-saosac.cgi).  We then
fit the ChaRT PSF in order to find an analytic description of it.
Next we fit the total surface brightness profile with both the
analytic PSF and either one or two $\beta$--models
\citep[e.g.,][]{cav76} in order to describe the extended emission.
Finally, we used an F--test to assess the significance of the PSF by
comparing the PSF plus $\beta$--model to a $\beta$--model alone.

The results of the spatial analysis are summarized in
Table~\ref{tbl-3}.  All the objects, except 1246--41, required a PSF
at high significance level.  We found that only 0034--01 required no
$\beta$--model; it was fit well by the PSF plus the
background.  The object 0915--11 was fit well by a PSF plus
$\beta$--model out to $\approx$25 arcseconds.  Past $\approx$25
arcseconds the source appears to require another $\beta$--model, but
we could not get good statistics on a second $\beta$--model---perhaps
because the field of view was not large enough.  We found that
1246--41 does not require a PSF, and while it too appears to require
an additional $\beta$--model past $\approx$25 arcseconds, we could not
get good statistics on a second $\beta$--model.  For 1333--33 it was
required to fix the value for the core radius in order to arrive at
physically meaningful results---several different values were tried,
with the quoted value yielding the best fit.

\subsection{Timing Analysis}

We produced background subtracted light curves for each of the objects
in our sample. For 1216+06 we refer to the detailed study by 
Gliozzi, Sambruna, \& Brandt (2003).  We chose to produce light curves based on
\xmm\ data when possible, but when only \chandra\ or \sax\ data were
available, those data were used.  According to a $\chi ^2$ test for
constancy, we found there to be no statistically significant nuclear
variability in any of the objects on the time scales probed by our
observations, $\sim$ few hours.  This is likely due to
(1) the intrinsic weakness of the sources, which translates into a low
count--rate especially for
\chandra\ and \sax\ observations, and (2) the short exposures.

\subsection{Spectral Analysis}

Previous X--ray spectral studies of low power radio galaxies suggest
that at least two components are required to fit a spectrum: one, a
thermal component associated with the extended emission; two, a
power--law component associated with the unresolved nuclear point
source \citep{wor94}.  This is consistent with the results of our
spatial analysis, in which four out of the six sources required a
point--like source in diffuse emission.  With these considerations in
mind we began our spectral modeling with a power law, which dominates
at hard energies, and a soft component modeled as emission from a
collisionally--ionized plasma \citep[{\tt apec} in {\tt
XSPEC};][]{smi01}.  All of the spectral models also assume a column
density fixed to the Galactic level appropriate to the source
\citep{dic90}.  Additional components, such as intrinsic absorption,
covering fraction, or additional thermal components, were added to the
model when the data required them.

The results of the spectral analysis are summarized in
Table~\ref{tbl-4}.  When a source was observed with both \xmm\ and
\chandra\, we quote the spectral model derived from the \xmm\ data,
since it was more statistically sound.  Seven out of the nine objects
require a power law with $\Gamma$ between $\sim 1.5$ and $\sim 2.7$,
with the average value being $\sim 1.9$.  The extremely flat value of
$\Gamma$ for 1251--12 might be related to the poor photon statistics.
Six out of the nine objects required at least one thermal component.
The temperatures show a bimodal distribution around 0.6 keV and $>1.5$
keV, which probably reflects the different environments---poor and
rich clusters respectively.  Five out of the nine objects required
intrinsic absorption with values ranging from $\sim 10^{21}$ to $\sim
10^{23}$ \nh, with the average value being $\sim 4 \times 10^{23}$.
We tried to add a \feka\ line at 6.4 keV, but this was not
statistically required for any of the objects except 1216+06
\citep[see][]{sam03}.

The spectral results are in good agreement with the findings of the
spatial analysis.  The surface brightness profile of 0034--01 was
found to not require a $\beta$--model, indicating that little to no
extended emission is present; indeed, we see that no thermal component
is required to fit the object's spectrum.  We also found that the
surface brightness profile of 1246--41 did not require a PSF,
indicating a heavily absorbed and/or very weak point source; this too
conforms to the spectral results, in that no power law was required
to fit the object's spectrum.

\section{Discussion} 

This paper seeks to shed light on the role of WLRGs/LINERs in the
context of uncovering the relationship between ``normal'' galaxies and
powerful AGN.  The results of the X--ray analysis allow us to address
several important questions towards this end: (1) What is the role of
obscuration in WLRGs/LINERs? (2) What is the nature of accretion in
such objects? (3) How do the X--ray properties of WLRGs/LINERs compare
to those of other classes of radio galaxies?

\subsection{The role of obscuration}

According to the unified model of AGN \citep[e.g.,][]{urr95}, the
central regions of AGN appear to contain a molecular torus that
prevents the penetration of radiation from the nucleus along certain
lines of sight.  This can cause intrinsically similar AGN to look
remarkably different from different viewing angles.  Recently,
however, the presence of an obscuring torus in low--power radio
galaxies, specifically FRIs, has been called into question
\citep[henceforth C99C02]{chi99,cap02}, based on \hst\ observations of
the 3C and B2 catalogs.  Since our sample is composed of low--power
radio galaxies, and because our analysis is focused on the X--rays, we
are in a position to contribute to this debate.  X--rays, due to their
high penetrating power, are the ideal means by which the presence of
an obscuring torus can be investigated.

The results of our spectral analysis show that 4 of the 9 objects in
our sample require an absorption component on the order of $10^{22}$
\nh, which qualifies as significant absorption.  Out of those 4
objects, 3 are confirmed LINERs, while 1 is classified as conflicting.
In addition to these 4 objects there are 2 others that, while not
heavily absorbed, require an absorption component on the order of
$10^{21}$ \nh; one of these additional objects is a confirmed LINER,
the other is a possible LINER.
Thus, 6 of 9 objects in our sample show signs of absorption, which
suggests that some low--power radio galaxies may indeed house
obscuring tori.  Moreover, all 6 of these absorbed objects are
classified as FRIs or FRI/FRIIs , which is in 
contrast to the hypothesis that low--power radio galaxies, specifically 
FRIs, lack a molecular torus (C99C02).  The apparently surprising result 
that the only
FRII in our sample (1637--77) is not absorbed may be explained if the object
can be classified as a Low Excitation Galaxy (LEG); Chiaberge, Capetti, \& 
Celotti (2002) explain that, while LEGs may be classified morphologically as 
FRIIs, their radio to optical nuclear properties are indistinguishable from 
FRIs.  The [O$_{~\rm{II}}$]$/$[O$_{~\rm{III}}$] ratio for 1637--77 places it
right on the border of the LEG classification, but if 1637--77 can be 
classified as an LEG, we should expect it to be unabsorbed.

In order to address the apparent discrepancy between the optical/UV and 
the X--ray absorption properties, one must keep in mind that the inner 
regions of the accretion flow ($\sim10-50~{\rm{R_G}}$), where most of the
radiation is supposed to originate, is inaccessible to any observatory.  
Even though the angular resolution of \hst\ ($\sim0.1$ arcseconds) is much
better than \chandra\ ($\sim1.0$ arcseconds) and \xmm\ ($\sim10$ arcseconds),
the region probed by \hst\ is not closer to the black hole than the X--ray 
satellites.  Indeed, assuming a typical black hole mass of $5\times10^8~{M_
{\odot}}$ at a distance of $40~\rm{Mpc}$ (corresponding to $z\sim0.01$), the 
inner $50~{\rm{R_G}}$ subtends a region of angular size $\sim10^{-4}$ 
arcseconds, which is three orders of magnitude smaller than the spatial 
resolution of \hst.  This indicates that the unresolved X--ray and optical 
sources simply reflect the PSF of the respective instrument rather than the 
physical size of the emitting regions.  

In light of these considerations, it is entirely possible that the X--ray
emission is produced in a region closer to the black hole than that of the
optical/UV emitting region.  Indeed, theoretical arguments predict that 
the location of the X--ray emitting region is closer to the black hole in 
both accretion and jet models.  This hypothesis is supported by the difference
in absorption properties found in the optical/UV and the X-rays, if it is
assumed that two distinct media are responsible for the absorption 
\citep{wei02,sam03,lew03}.  In this theory the optical/UV absorber is
a dusty medium lying farther away from the nucleus than the X--ray
absorber, which is a neutral or ionized gas largely free of dust.  Further
support for the hypothesis that the X--rays are produced in a region interior
to the optical/UV emitting region comes from the recent work of Donato, 
Sambruna, \& Gliozzi
(2004).  Those authors find that a substantial fraction of the compact cores
detected in the optical \citep{chi99}, do not have an unresolved X--ray
counterpart, which suggests a different origin and/or nature for optical/UV 
and X-ray radiation.  Additionally, Donato et al. show that neither X--ray 
obscuration nor the detection of unresolved X--ray cores is related to the 
presence of dust lanes, which are responsible for the optical absorption.  
Finally, we note that there is no discrepancy between
this paper and Donato et al. (2004) concerning obscuration; in that paper all 
the FRIs except the LINER 1216+06 (NGC 4261) show negligible absorption 
(N$_H<10^{21}~\rm{cm^{-2}}$).  This suggests that the absorption mechanisms 
in WLRGs/LINERs may be different than those in other low--luminosity radio 
galaxies.

\subsection{The nature of accretion}

The origin of X--rays in radio galaxies is a topic of considerable
debate.  Specifically the question is: Do X--rays have their origin in
the process of accretion or at the unresolved base of the jet?  The
correlation between the radio and X--ray core luminosities in low and
high luminosity radio galaxies
\citep[e.g.,][]{fab84,can99,wor94,har98} has often been used as an
argument in favor of a common origin at the base of the jet.  However,
this correlation has recently been shown by Merloni, Sebastian, \& Di Mateo 
(2003) to
derive from a more general correlation involving both the X--ray
luminosity and the black hole mass---the so--called ``fundamental
plane''---in which the radio is related to the jet, and the X--rays
are related to inefficient accretion onto the black hole.

This idea is supported by the detailed study of 1216+06 (NGC 4261) by
Gliozzi et al. (2003)---their argument is summarized by three main
points: (1) they found there to be a high Bondi accretion rate along
with a low AGN luminosity, suggesting the presence of a radiatively
inefficient accretion flow; (2) they observed an Fe line at $\approx7$
keV of nuclear origin; (3) they studied the temporal and spectral
variability, and combined this with independent information on jet
energetics, which lead them to conclude that the bulk of the X--ray
emission originates from a radiatively inefficient accretion flow,
with negligible jet contribution.  Based on the similarity between
1216+06 and the other objects in our sample, we explored the
possibility that the X--rays for all the objects in our sample are
related to accretion and not the base of the jet.

\begin{figure}
\begin{center}
\includegraphics[width=8cm]{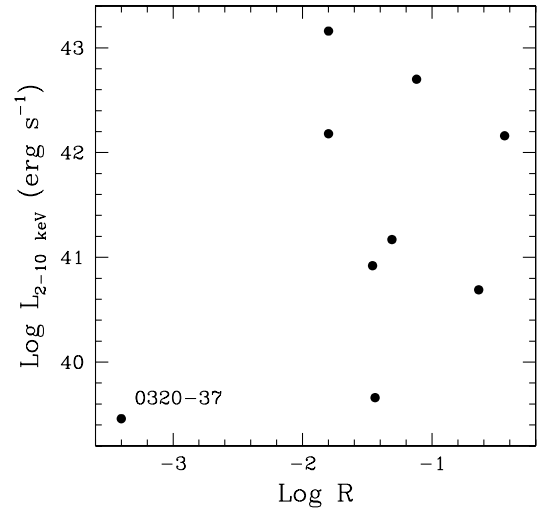}
\end{center}
\caption{Plot of radio core dominance R vs. the 2--10 keV intrinsic luminosity.  The parameter R 
is defined as a ratio of the core-to-lobes 5 GHz luminosities  from Morganti et al. (1993).
No correlation between the core X--ray luminosity and the radio core dominance is apparent.}
\label{isoemission}
\end{figure}

One way to explore the origin of nuclear emission is via the
correlation (or lack thereof) with the radio core dominance
($R=L_{core}/L_{lobe}$)---this has been done in the optical by Kharb
\& Shastri (2004).  By demonstrating a strong correlation between
optical point--like emission and radio core dominance, Kharb \&
Shastri found the optical emission to be strongly beamed, and thus
produced by the jet, confirming the results of Chiaberge, Capetti, 
\& Celotti (1999).
We adopted this approach for our X--ray sample
(see Fig.~\ref{isoemission}, Table~\ref{tbl-6}), but found there to 
be no correlation
between the core X--ray emission and the radio core dominance,
indicating that the emission is not beamed, but rather is isotropic
and thus likely to be related to accretion.  On the other hand, 
there is some indication that in non--LINER FRIs at least a fraction of the 
X--ray emission is beamed, and thus related to the base of the jet 
\citep{donprep}.

Many of the objects in our sample require a powerlaw component for
their spectral fit, but the powerlaw components (ranging from
$\Gamma\sim1.5$ to $\sim2.1$, excluding the \sax\ observation with
$\Gamma\sim2.7$) can be a good representation for both jet dominated
and accretion dominated emission.  The lack of a strong \feka\ line
suggests the lack of a standard accretion disk; however, both the
radiatively inefficient accretion and the jet scenario can account for
this result.  And while temporal analysis can yield insights into the
nature of accretion, our results yield no such clues.

More insightful information comes from the calculations of the
bolometric and Eddington luminosities.  We used estimates of the black
hole masses based on stellar velocity dispersions to calculate
$L_{{\rm Edd}}$, and the relationship $0.1L_{{\rm bol}}=L_{\rm X}$ in
order to calculate $L_{{\rm bol}}$.  For Seyfert--like objects, which
are thought to have a standard thin--disk, the ratio of $L_{{\rm
bol}}$ to $L_{{\rm Edd}}$ is $\sim10-20\%$; an Eddington ratio much
less than 10\% could signify inefficient accretion.  Some accretion
properties are listed in Table~\ref{tbl-5}.  In our sample we find
that the majority of sources have values of $L_{{\rm bol}}/L_{{\rm
Edd}}$ between $10^{-4}$ and $10^{-7}$.  Even the \xmm\ observation of
0915--11, with its high X--ray luminosity, has $L_{{\rm bol}}/L_{{\rm
Edd}}\sim10^{-3}$.  This is a model--independent indication of
Radiatively Inefficient Accretion Flow (RIAF).  Using the results of
our spatial analysis, specifically the gas density profile obtained by the 
de--projection of the surface brightness profile, we calculated 
\.M$_{{\rm Bondi}}$, which serves
as a rough estimate of accretion onto the black hole.  With {\it
\.M}$_{{\rm Bondi}}$ and $L_{{\rm bol}}$ we can use the formula
$L_{{\rm bol}}=\eta${\it \.M}$_{{\rm Bondi}}c^2$ to make an estimate
of the radiative efficiency of accretion $\eta$.  We find values on
the order of $10^{-5}-10^{-6}$, except for the object 0915--11, which
using the \chandra\ observation yields $\eta=0.01$, and using the
\xmm\ observation yields $\eta=0.09$.  Those objects with $\eta<<0.1$
are very likely to have inefficient accretion, but even 0915--11,
based on the \chandra\ observation, is a candidate.  However, the
\xmm\ observation makes the accretion scenario for 0915--11 ambiguous.

\subsection{Comparison with previous results}

As mentioned above, Donato et al. (2004) have analyzed a larger sample of FRIs 
from the B2 and 3C catalogues.  Those authors found there to be no absorption
of X--rays in FRIs, except in the LINER 1216+06 (NGC 4261), which is consistent
with our results.  While Donato et al. found that the majority of the X--rays 
in classical FRIs are related to inefficient accretion onto the black hole, 
their analysis suggests that at least a fraction of the X--rays are beamed 
and thus potentially related to the base of the jet.  This lends further 
support to the idea that WLRGs/LINERs form a distinct group compared to 
classical FRIs.

\begin{figure}
\begin{center}

\includegraphics[width=8.3cm]{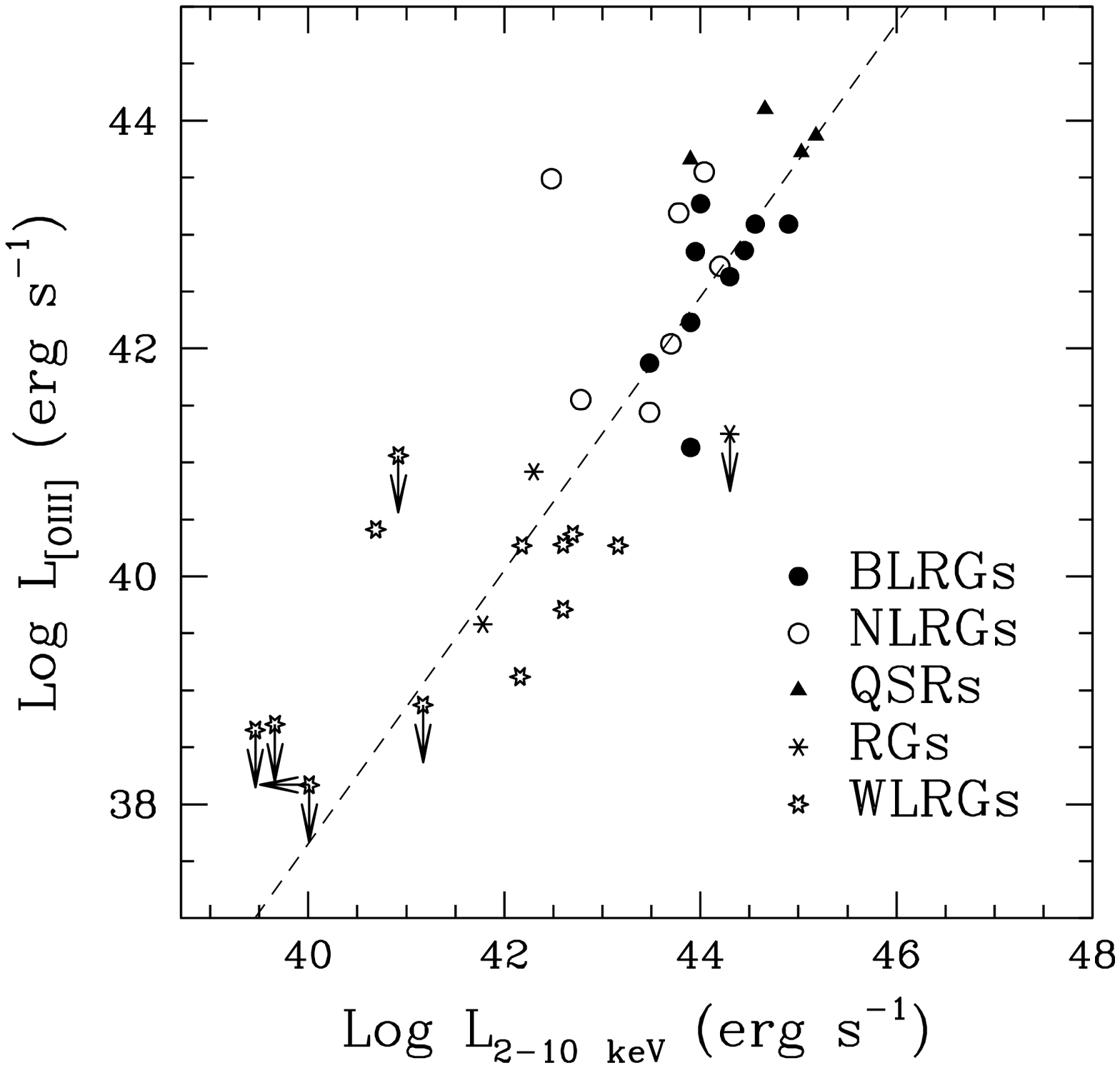}\includegraphics[width=8.3cm]{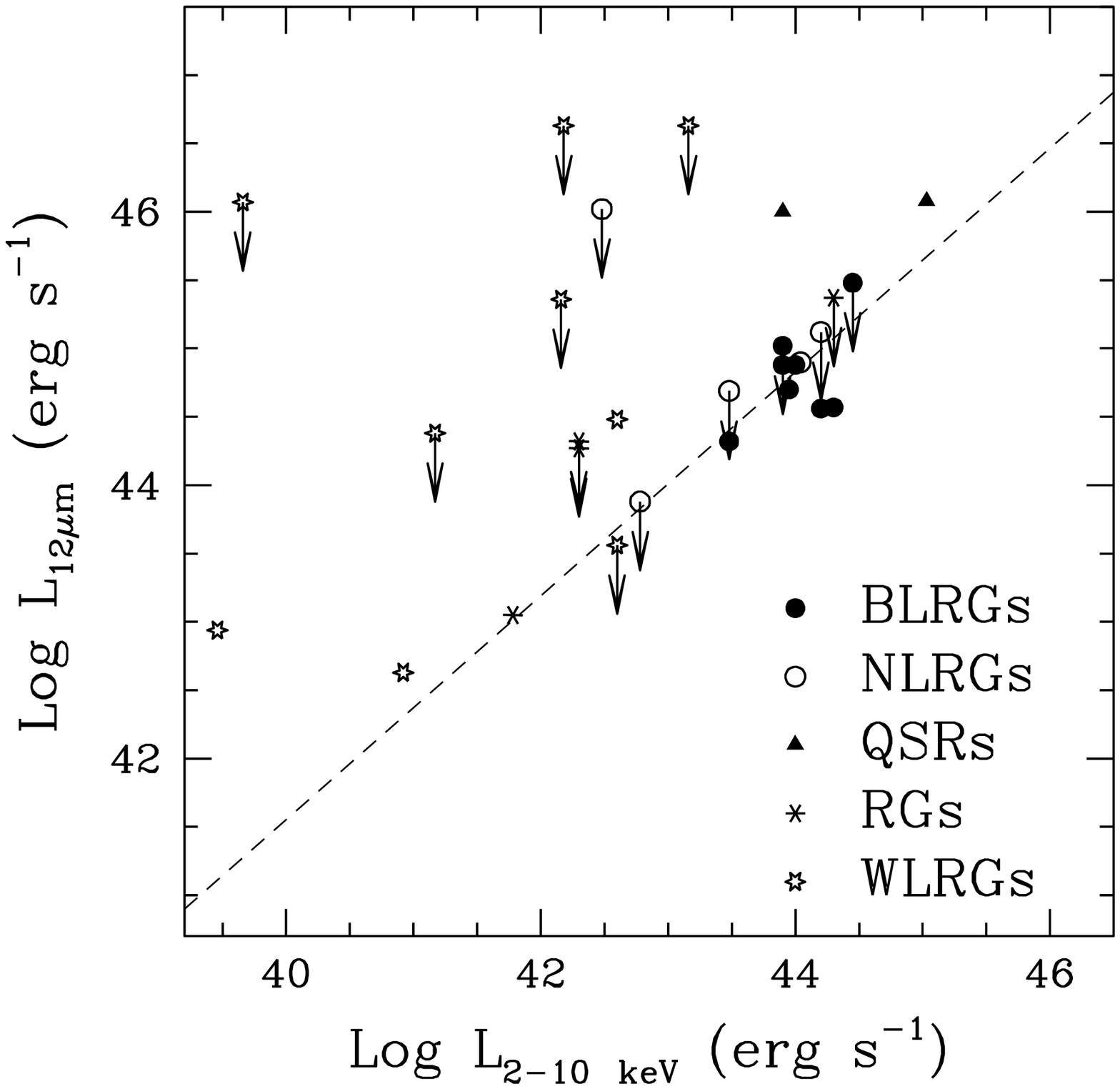}

\includegraphics[width=8.3cm]{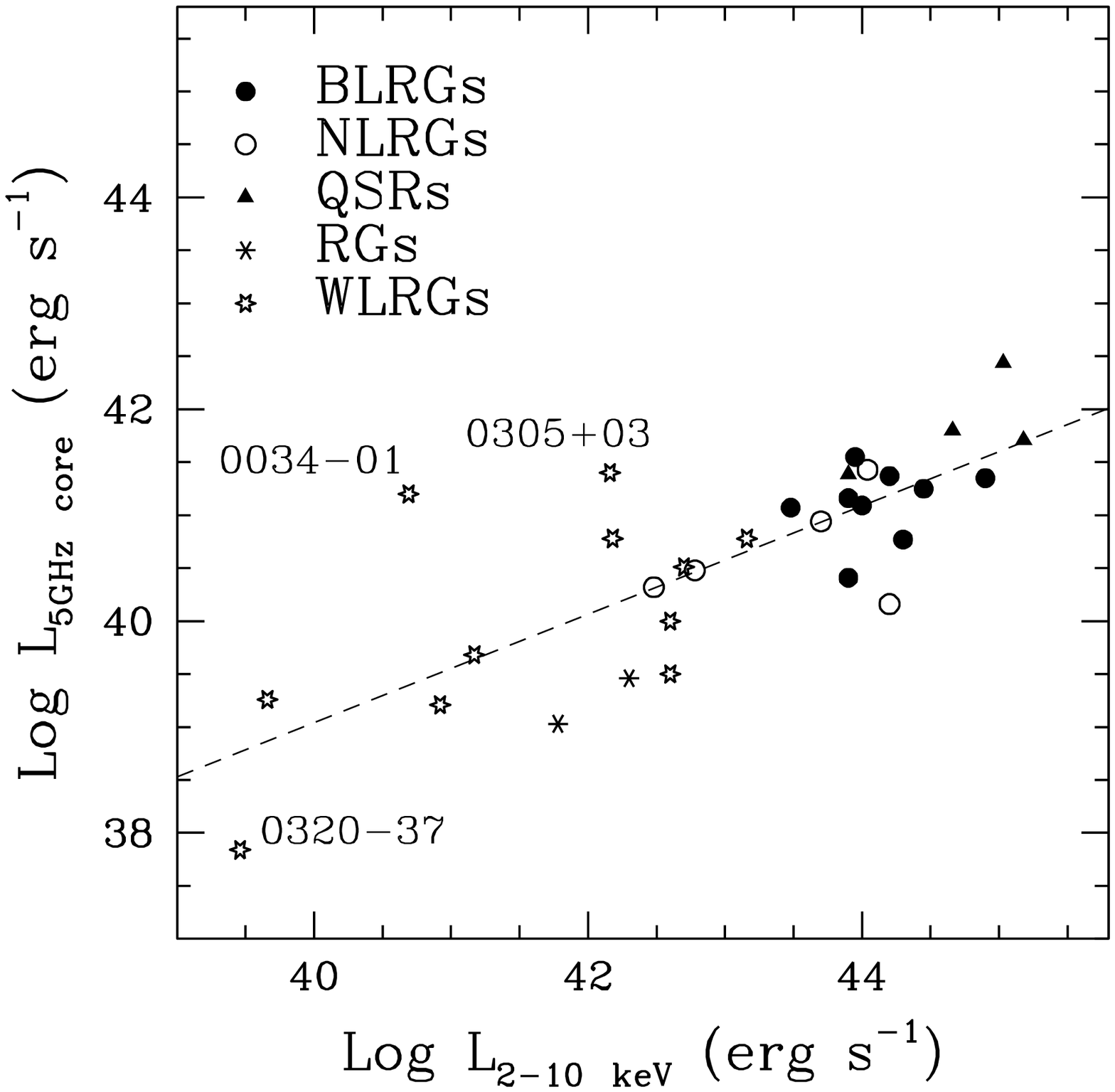}\includegraphics[width=8.3cm]{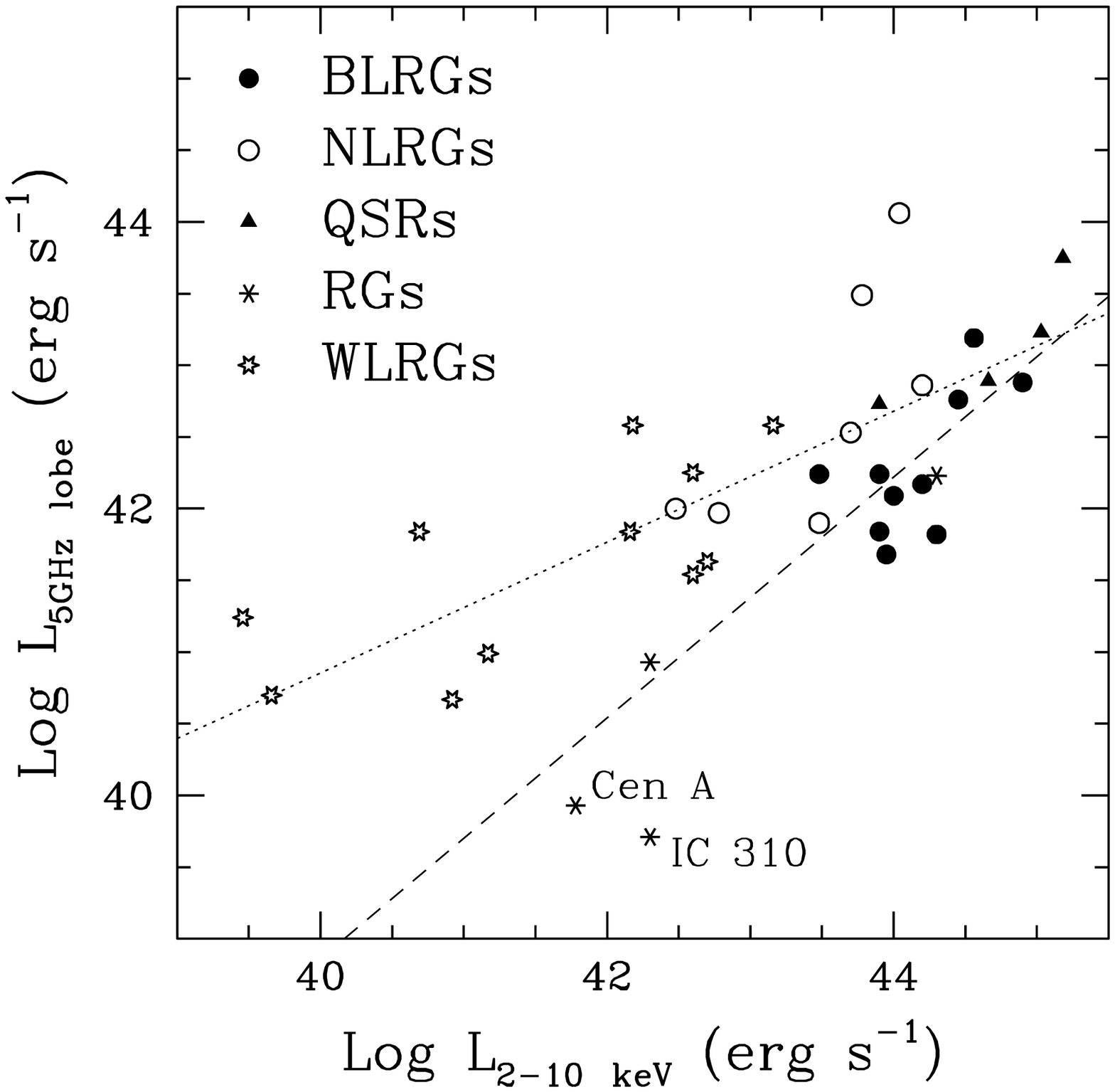}

\end{center}
\caption{The top--left figure shows the correlation between the 
luminosity of the [O$_{~\rm{III}}$]$\lambda$5007 emission line and 
the intrinsic 2--10 keV luminosity for various subclasses of 
radio--loud AGN.  Excluding the upper limits, a least squares fit yields the relationship 
$\log L_{[{\rm O_{III}}]}=1.2~\log L_{\rm 2-10~keV}-10.33$.  
The top--right figure shows the correlation between the MIR 
emission at 12$\mu m$ and the intrinsic 2--10 keV luminosity.  
Excluding the upper limits, the correlation is consistent with 
Sambruna et al. (1999).  The bottom--left figure shows the core 
radio power at 5 GHz and the 
intrinsic 2--10 keV luminosity; a least 
squares fit yields the relationship 
$\log L_{\rm core}=0.51~\log L_{\rm 2-10~keV}+18.56$.
The bottom--right figure shows the 
correlation between the lobe radio power at 5 GHz and the 
intrinsic 2--10 keV luminosity.  A least 
squares fit of the BLRGs, NLRGs, QSRs, and RGs yields the 
relationship $\log L_{\rm lobe}=0.84~\log L_{\rm 2-10~keV}+5.26$ 
(dashed line).  However, the WLRGs formed a distinct group; a least 
squared fit yields the relationship 
$\log L_{\rm lobe}=0.46~\log L_{\rm 2-10~keV}+22.6$ 
(dotted line).
\label{correlations}}
\end{figure}

A more general X--ray analysis of radio--loud AGN was made by Sambruna, 
Eracleous, \& Mushotzky (1999) using \asca.  In that paper the authors 
found there to
be correlations for the [O$_{~\rm{III}}$]$\lambda$5007, the 5 GHz
lobe, and the 12$\rm \mu m$ luminosities to the 2--10 keV intrinsic 
X--ray luminosity.  Those correlations demonstrated a slight lobe 
radio excess in
WLRGs.  We have updated the plots of Sambruna et al. to include those
WLRGs that were not in that paper, and have introduced a plot of 
the 5 GHz core luminosity and the 2--10 keV intrinsic X--ray
luminosity (see Table~\ref{tbl-6}, and
Fig.~\ref{correlations}).  These updated plots show that the WLRGs do
indeed demonstrate a lobe radio excess; moreover, they form a distinct
class as compared to the other subclasses of radio--loud AGN.  To
quantify the degree of linear correlation between the 2--10 keV
intrinsic luminosity and the 5 GHz lobe 
radio power for the WLRGs (and
the other correlations that follow), we calculated the linear
correlation coefficient $r$ and computed the chance probability
$P_c(r;N)$ that a random sample of $N$ uncorrelated pairs of
measurements would yield a linear correlation coefficient equal to or
greater than $|r|$.  We also calculated the Spearman and Kendall parameters, 
which confirm the $P_c(r;N)$ results. We found the probability of 
correlation between
$L_{\rm lobe}$ and $L_{\rm 2-10~keV}$ for the WLRGs to be
$\approx98\%$, with a least squares fit yielding the relationship
$\log L_{\rm lobe}=0.46~\log L_{\rm 2-10~keV}+22.6$ 
($0.46\pm0.1$, $22.6\pm5$), represented by the dotted line in
Fig.~\ref{correlations}.  The remaining subclasses of AGN formed a
distinct group of their own with a probability of correlation
$>99.9\%$; however, the least squares fit of this group yielded a
significantly different relationship: $\log L_{\rm lobe}=0.84~\log
L_{\rm 2-10~keV}+5.26$ ($0.84\pm0.1$, $5.26\pm6$),
represented by the dashed line in Fig.~\ref{correlations}.  This 
lobe excess in the WLRGs raises the possibility that there once was
much more intense jet activity in these objects than is shown today.

The plot of the core radio power vs. the intrinsic 2--10 keV 
luminosity shows that the WLRGs follow the correlation of the 
other subclasses of radio--loud AGN.  The probability of correlation 
is $>99.9\%$, and a least squares fit yields the relationship
$\log L_{\rm core}=0.51~\log L_{\rm 2-10~keV}+18.56$ 
($0.51\pm0.08$, $18.56\pm3.3$).

The plot
of the MIR luminosity vs. the intrinsic 2--10 keV luminosity suggests
that there may also be an MIR excess.  However, the MIR data for WLRGs
are largely upper limits, so this suggestion should be taken with
caution; also, because there were so many upper limits, we were not
able to better constrain the correlation found by Sambruna et
al. (1999), thus Fig.~\ref{correlations} plots that previously found
relationship: $L_{\rm 12\mu m}=0.83~\log L_{\rm 2-10~keV}+8.28$.
Finally, we found the probability of correlation between $L_{[{\rm
O_{III}}]}$ and $L_{\rm 2-10~keV}$ to be $>99\%$, with a least squares
fit yielding the relationship $\log L_{[{\rm O_{III}}]}=1.2~\log
L_{\rm 2-10~keV}-10.33$ ($1.2\pm0.2$, $-10.33\pm7$),
which is consistent within the errors with Sambruna et al. (1999).  
We notice that the WLRGs tend be have a slight [O$_{~\rm{III}}$] deficiency; 
however, due to the small number of data points, we could not get good 
statistics for a least squares fit of the WLRGs as a distinct class.

\section{Summary}

This paper shows the results of an X--ray analysis of 9
WLRGs/LINERs---the main findings are summarized as follows:

(1) We found that 6 out of the 9 objects in our sample are absorbed in
the X--rays.  Four of these 6 objects are absorbed on the order of
$10^{22}$ \nh, which qualifies as significant absorption.  The other 2
objects show slightly lower values of absorption ($10^{21}$ \nh).
These results, combined with those of a larger sample of non--LINER 
FRIs, for which there is a systematic lack of absorption, suggests that 
WLRGs/LINERs form a distinct class compared to other low--power radio
galaxies.

(2) The X--rays in this sample of WLRGs/LINERs appear to be related to
accretion.  The object 1216+06 has been carefully studied by Gliozzi
et al. (2003) and the X--rays of that object are found to be related
to accretion.  We found that the rest of the objects in our sample
behave like 1216+06, and while we can not rule out that a fraction of
the X--rays come from the base of the jet, we performed a model
independent test and found that the X--rays in this sample do not
appear to be beamed.  On the other hand, the recent analysis of a sample
of classical FRIs reveals that, while the majority of the X--rays are
related to inefficient accretion onto the black hole, a fraction of the
X--rays may be beamed and thus likely related to the base of the jet.
This lends further 
support to the idea that WLRGs/LINERs form a distinct class compared to 
other low--power radio galaxies.

(3) Assuming that all the X-ray emission is related to accretion, our
tests concerning the nature of accretion are in favor of it being
radiatively inefficient.  But even if some of the X--rays do stem from
the base of the jet, the argument for radiatively inefficient
accretion is only strengthened---the fraction of X--rays related to
accretion would be even lower than we assumed them to be, which in
turn would yield even low efficiency.

Our results are based on a small sample of WLRGs/LINERs.  These
results should be confirmed by an analysis of a larger sample, which
we plan to undertake in a future paper. 

\vskip 0.5 cm

\begin{acknowledgements}

We gratefully acknowledge the financial support provided by
Smithsonian grant GO3--4123A (ASR), and NASA LTSA grant NAG5--10708
(RMS, MG). RMS gratefully acknowledges support from an NSF CAREER
award and from the Clare Boothe Luce Program of the Henry Luce
Foundation.

\end{acknowledgements}

\clearpage

\input{tab1}
\input{tab2}
\input{tab3}
\input{tab4}
\input{tab5}

\input{tab6}

\clearpage

\clearpage

\clearpage

\clearpage

\end{document}

%% file: tab1.tex
\begin{table}
\begin{center}
\caption{Object Sample \label{tbl-1}}
\begin{tabular}{ccccccccc}
\\
\tableline\tableline
Object & Alt. Name & RA & Dec & $z$ & N$_{H}^{Gal}$ & log M$_{BH}$ & FR & LINER
\\
(1) & (2) & (3) & (4) & (5) & (6) & (7) & (8) & (9)\\
\hline  
\\
0034$-$01 & 3C 15 & 00 37 04.1 & $-$01 09 08 & 0.007 & 2.9 & 8.81 & I/II & C \\
0305+03 & 3C 78 & 03 08 26.2 & +04 06 39 & 0.029 & 10.3 & 8.71 &I & PL \\
0320$-$37 & For A & 03 22 41.7 & $-$37 12 30 & 0.006 & 1.9 & 8.31 &I & PL \\
0915$-$11 & Hyd A & 09 18 05.7 & $-$12 05 44 & 0.054 & 4.9 & 8.97 &I & L\\
1216+06 & 3C 270 & 12 16 23.2 & +05 49 31 & 0.007 & 1.5 & 8.72 &I & L \\
1246$-$41 & NGC 4696 & 12 48 49.3 & $-$41 18 40 & 0.010 & 8.1 & 8.65 & I & L \\
1251$-$12 & 3C 278 & 12 54 36.1 & $-$12 33 48 & 0.015 & 3.6 & 8.64 &I & L \\
1333$-$33 & IC 4296 & 13 36 39.0 & $-$33 57 57 & 0.013 & 4.1 & 9.16 &I/II & L\\
1637$-$77 & PKS 1637$-$77 & 16 44 16.1 & $-$77 15 48 & 0.043 & 8.7 & 8.68 & II & L\\
\tableline
\end{tabular}
\tablecomments{{\bf Columns:} (1) object's IAU name; (2) object's common name; (3)-(4) right ascension and declination (Equ J2000) from NED; (5) redshift from NED; (6) Galactic equivalent hydrogen column (in units of ${10}^{20}$ \nh); (7) black hole masses in units of $\log M_{\odot}$ calculated using the velocity dispersion--$M_{BH}$ correlation (Ferrarese 2000), except 0034--01 and 1637--77 from Bettoni et al. 2003, and 1216+06 from Ferrarese et al. 1996; (8) Fanaroff \& Riley radio classification from Tadhunter et al. 1998, except for 0034$-$01 which is from Leahy et al. (1997); (9) classification from Lewis et al. (2003)---L=LINER, PL=Possible LINER, C=Conflicting.}

\end{center}
\end{table}

\clearpage

%% file: tab2.tex
\begin{table}
\begin{center}
\caption{Observation Log \label{tbl-2}}
\begin{tabular}{cccccccc}
\\
\tableline\tableline
Object & Sat & Date & Exp & r$_{ext}$ & Counts
\\
(1) & (2) & (3) & (4) & (5) & (6) \\
\hline  
\\
0034$-$01 & CXO & 00--11--06 & 25.4 & 1.5 & 493$\pm22$\\
0305+03 & SAX & 97--01--07 & 19.4 & 240 & 672$\pm27$\\
0320$-$37 & CXO & 01--04--17 & 29.1 & 1.5 & 464$\pm22$\\
0915$-$11 & CXO & 99--11--02 & 17.8 & 1.5 & 400$\pm20$\\
 & XMM & 00--12--08 & 20.2 & 15.0 & 28810$\pm170$\\
1216+06 & CXO & 00--05--06 & 32.0 & 1.5 & 1738$\pm41$\\
 & XMM & 01--12--16 & 21.3 & 20.0 & 6454$\pm85$\\
1246$-$41 & CXO & 00--05--22 & 31.7 & 12.0$^{\bf a}$ & 16719$\pm136$\\
 & XMM & 02--01--03 & 40.0 & 15.0 & 33055$\pm236$\\
1251$-$12 & CXO & 02--06--16 & 48.6 & 1.5 & 154$\pm12$ \\
1333$-$33 & CXO & 01--12--15 & 24.3 & 1.5 & 1723$\pm42$\\
1637$-$77 & XMM & 02--02--23 & 3.5 & 15.0 & 1961$\pm45$\\
\tableline
\end{tabular}

\tablecomments{{\bf Columns:} (1) object's name; (2) satellite with which the object was observed---CXO=\chandra, XMM=\xmm, SAX=\sax; (3) date of observation (yy--mm--dd); (4) effective exposure time in kilo--seconds; (5) radius of circular extraction region centered about the nucleus in arcseconds; (6) background corrected number of counts in the extraction region---the energy range of the counts is depentent upon the detector used: ACIS-S=0.5--8.0 keV, XMM=0.3--10.0 keV, SAX=0.3--10.0 keV.}
\end{center}
\end{table}

\clearpage

%% file: tab3.tex
\begin{table}
\begin{center}
\caption{Results of Spatial Analysis \label{tbl-3}}
\begin{tabular}{cccccc}
\\
\tableline\tableline
Object & R$_c$ & $\beta$ & N$_{\beta}$ & P$_{PSF}$\\
(1) & (2) & (3) & (4) & (5)\\
\tableline
\\
0034$-$01 & - & - & - & $>$99.9 \\

0320$-$37 & 4.6$\pm0.3$ & 0.61$\pm0.06$ & $12.59\pm0.008 $& $>$99.9 \\

0915$-$11 & 15.9$\pm1.6$ & 0.47$\pm0.10$ & $15.85\pm0.316$ & $>$99.9 \\

1246$-$41 & 18.8$\pm2.0$ & 0.64$\pm0.20$ & $12.59\pm0.199$ & 16.6 \\

1251$-$12 & 1.8$\pm0.5$ & 0.49$\pm0.11$ & $2.512\pm0.631$ & $>$99.9 \\

1333$-$33 & 0.8$_{fixed}$ & 0.56$\pm0.01$ & $199.5\pm7.943$ & $>$99.9 \\
\tableline
\end{tabular}

\tablecomments{{\bf Columns:} (1) object's name; (2) core radius in arcseconds; (3) $\beta$ value; (4) normalization on the $\beta$--model in units of $10^{-4}$; (5) probability that a PSF is required according to an F--test.}
\end{center}
\end{table}

\clearpage

%% file: tab4.tex
\begin{table}
\begin{center}
\caption{Results of Spectral Analysis \label{tbl-4}}
\begin{tabular}{ccccccccc}
\\
\tableline\tableline
Object & model\tablenotemark{a} & kT & Abund & N$_{H}^{Add}$ & CF & EW & $\Gamma$ & $\chi^{2}_{red}$/d.o.f.\\
(1) & (2) & (3) & (4) & (5) & (6) & (7) & (8) & (9) \\
\tableline
\\
0034$-$01 & I & - & - & 8.8$^{+3.5}_{-2.4}$ & 0.86$\pm{0.06}$ & $<$170 & 1.58$^{+0.29}_{-0.32}$ & 0.73/19 \\

0305+03 & IV & - & - & - & - & $<$1900 & 2.74$^{+0.26}_{-0.25}$ & 0.98/31 \\

0320$-$37 & II & 0.6$\pm{0.1}$ & 0.2$_{fixed}$ & 0.3$^{+1.3}_{-0.3}$ & - & - & 1.78$^{+1.21}_{-0.40}$ & 0.90/15 \\

0915$-$11  & VI & 1.8$^{+0.34}_{-0.56}$& 0.2$^{+0.05}_{-0.09}$& 6.0$^{+6.3}_{-2.4}$& - & $<$51.3 & 2.14$^{+1.26}_{-0.58}$& 1.03/548 \\
 & & 4.3$^{+2.30}_{-1.01}$& 1.2$^{+0.00}_{-0.64}$& & & & \\

1216+06 & III & 0.7$^{+0.01}_{-0.02}$ & 1.0$_{fixed}$ & 5.1$^{+1.1}_{-1.2}$ & 0.81$^{+0.06}_{-0.09}$ & 230$^{+166}_{-134}$\tablenotemark{c} & 1.46$^{+0.17}_{-0.31}$ & 1.12/359\\

1246$-$41 & V & 0.7$\pm{0.01}$& 1.2$^{+0.00}_{-0.07}$& 0.1$\pm0.01$\tablenotemark{b} & - & $<$374 & - & 1.23/483\\
 &  & 1.5$\pm{0.02}$& 1.2$^{+0.00}_{-0.02}$& & & & \\

1251$-$12 & VII & 0.6$\pm{0.1}$ & 1.2$^{+0.00}_{-0.93}$ & - & - & - & -0.07$^{+0.58}_{-0.69}$ & 334/435\tablenotemark{d} \\

1333$-$33 & II & 0.6$\pm{0.3}$ & 1.0$_{fixed}$ & 1.1$^{+0.6}_{-0.5}$ & - & - & 1.52$^{+0.42}_{-0.37}$ & 1.22/66 \\

1637$-$77 & IV & -  & - & - & - & $<$375 & 1.90$\pm{0.06}$ & 0.76/88 \\

\tableline

\end{tabular}

\tablenotetext{a}{I=wabs(zpcfabs(powerlaw)); II=wabs(apec+zwabs(powerlaw)); III=wabs(apec+zpcfabs(powerlaw)); IV=wabs(powerlaw);  V=wabs(apec+apec); VI=wabs(apec+apec+wabs(powerlaw)); VII=wabs(apec+powerlaw).}
\tablenotetext{b}{This is not an additional absorption component; such a component did not help the fit.  This is the number to which the galactic absorption rose when we let it free.}
\tablenotetext{c}{More precise values of $E_{line}$ and $\sigma$ were able to be fit for this object: $E_{line}=6.99^{+0.08}_{-0.09}$ keV and $\sigma=0.03^{+0.15}_{-0.03}$ keV.}
\tablenotetext{d}{Due to the very low number of counts, we used the C--statistics to fit this model.  Column (9) therefore gives the C--statistics and the number of pha bins.}
\tablecomments{{\bf Columns:} (1) object's name; (2)
spectral model (see note); (3) plasma temperature in keV; (4) metal
abundances (solar=1.0); (5) Absorption column density at the redshift
of the source in units of $10^{22}$ \nh; (6) covering fraction; (7)
equivalent width in eV at $E_{line}=6.4$ keV with the physical width
of the Gaussian model 
fixed at $\sigma=0.01$ keV---see note for 1216+06; (8) photon Index; 
(9) reduced $\chi^2$ and degrees of freedom.}
\end{center}
\end{table}

\clearpage

%% file: tab5.tex
\begin{table}
\begin{center}
\caption{Accretion Properties \label{tbl-5}}
\begin{tabular}{cccc}
\\
\tableline\tableline
Object & ${\rm L_{{\rm bol}}/L_{{\rm Edd}}}$ & {\it \.M}$_{\rm Bondi}$
& $\eta$
\\
(1) & (2) & (3) & (4) \\
\tableline
\\
0034$-$01          & 6.020 & -     & -     \\
0305+03            & 219.5 & -     & -     \\
0320$-$37          & 1.126 & 0.05 & 1.106 \\
0915$-$11          & 132.2 & 0.03 & 1017  \\
0915$-$11\tablenotemark{a} & 1234  & 0.03 & 9490  \\
1216+06            & 19.20 & 0.04 & 47.28 \\
1246$-$41          & 1.825 & 0.07 & 2.485 \\
1251$-$12          & 0.853 & 0.05 & 1.633 \\
1333$-$33          & 8.154 & 6.41 & 0.407 \\
1637$-$77          & 834.3 & -     & -     \\
\tableline
\end{tabular}

\tablenotetext{a}{Refers to values based on {$\rm L_{2-10 kev}$} from \xmm\ observation.}
\tablecomments{{\bf Columns:} (1) object's name; (2) ratio of the bolometric luminosity to the Eddington luminosity in units of $10^{-6}$; (3) Bondi accretion rate in units of M$_{\odot}~{\rm yr^{-1}}$; (4) radiative efficiency in units of $10^{-5}$.}
\end{center}
\end{table}

\clearpage

%% file: tab6.tex
\begin{table}
\begin{center}
\caption{Luminosities \label{tbl-6}}
\begin{tabular}{cccccc}
\\
\tableline\tableline
Object & L$_{\rm 5GHz}$ & L$_{\rm 12_{\mu m}}$ & L$_{\rm [O_{~\rm{III}}]}$ & L$_{\rm 2-10keV}$ & R
\\
(1) & (2) & (3) & (4) & (5) & (6) \\
\tableline
\\
0034$-$01 & 41.84 &          & 40.41    & 40.69 & $-$0.64\\
0305+03   & 41.84 & $<$45.36 & 39.12    & 42.16\tablenotemark{a} & $-$0.44 \\
0320$-$37 & 41.24 & 42.94    & $<$38.65 & 39.46 & $-$3.40            \\
0915$-$11 & 42.58 & $<$46.63 & 40.27    & 42.18 & $-$1.80            \\
          &   -   &   -      &   -      & 43.16\tablenotemark{b}& -  \\
1216+06   & 40.67 & 42.63    & $<$41.06 & 40.92\tablenotemark{b} & $-$1.46   \\
1246$-$41 &       &          & $<$38.17 & $<$40.01\tablenotemark{b}& \\
1251$-$12 & 40.70 & $<$46.07 & $<$38.70 & 39.66 & $-$1.44       \\
1333$-$33 & 40.99 & $<$44.38 & $<$38.87 & 41.17 & $-$1.31      \\
1637$-$77 & 41.63 &          & 40.37    & 42.70\tablenotemark{b} & $-$1.12 \\
\tableline
\end{tabular}
\tablenotetext{a}{Refers to \sax.}
\tablenotetext{b}{Refers to \xmm.}
\tablecomments{{\bf Columns:} Values in (2)--(5) are $log(\nu L{_\nu})$ (\lum).  References are as stated below except for 0320--37, 1216+06, and 1333--33 for which the values in columns (2)--(4), (6) come from Sambruna et al. (1999).  Column (2) lists the 5 GHz lobe luminosity (L$_{\rm tot}-$L$_{\rm core}$) from Morganti et al. (1993); column (3) lists the 12$_{\mu m}$ luminosity from Golombek et al. (1988); column (4) lists the [O$_{~\rm{III}}$]$\lambda$5007 line luminosity from Tadhunter et al. (1998); column (5) lists the 2--10 keV intrinsic luminosity, where the default intrument is \chandra; column (6) lists the R--value from Morganti et al. (1993), defined as the ratio of the core radio luminosity to the lobe radio luminosity at 5 GHz.}
\end{center}
\end{table}